\documentclass[twocolumn,aps,prba,superscriptaddress]{revtex4-1}
\usepackage[utf8]{inputenc}
\usepackage{amsmath}
\usepackage{amssymb}
\usepackage{graphicx}
\usepackage{booktabs}
\usepackage{physics}
\usepackage{siunitx}
\usepackage[format=plain]{caption}
\usepackage{subfig}
\usepackage[margin=0.9in]{geometry}
\usepackage{textcomp}
\usepackage{gensymb}
\usepackage{sidecap}
\usepackage[english]{babel}
\usepackage[autostyle]{csquotes}
\usepackage{braket}
\usepackage{array}
\usepackage{color}
\usepackage{multirow}

\begin{document}
\justifying
\captionsetup{justification=Justified,}

\title{Simulation of the crystallization kinetics of  Ge$_2$Sb$_2$Te$_5$ nanoconfined in superlattice geometries for phase change memories}
\author{Debdipto Acharya}
\affiliation{Department of Materials Science, University of Milano-Bicocca, Via R. Cozzi 55, I-20125 Milano, Italy}
\author{Omar Abou El Kheir}
\affiliation{Department of Materials Science, University of Milano-Bicocca, Via R. Cozzi 55, I-20125 Milano, Italy}
\author{Simone Marcorini}
\affiliation{Department of Materials Science, University of Milano-Bicocca, Via R. Cozzi 55, I-20125 Milano, Italy}

\author{Marco Bernasconi}
\affiliation{Department of Materials Science, University of Milano-Bicocca, Via R. Cozzi 55, I-20125 Milano, Italy}

\begin{abstract}
Phase change materials  are the most promising candidates for the realization of artificial synapsis for neuromorphic computing.  Different resistance levels corresponding to analogic values of the synapsis conductance can be achieved by modulating the size of an amorphous region  embedded in its crystalline matrix. Recently, it has been proposed that a superlattice made of alternating layers of the phase change compound Sb$_2$Te$_3$  and of the TiTe$_2$ confining material allows for a better control of multiple intermediate resistance states and for a lower drift with time of the electrical resistance of the amorphous phase. In this work, we  consider to substitute Sb$_2$Te$_3$ with the Ge$_2$Sb$_2$Te$_5$ prototypical phase change compound that should feature  better data retention. By exploiting molecular dynamics simulations with a machine learning interatomic potential, we have investigated the crystallization kinetics of Ge$_2$Sb$_2$Te$_5$ nanoconfined in geometries mimicking Ge$_2$Sb$_2$Te$_5$/TiTe$_2$ superlattices. It turns out that nanoconfinement induces a slight reduction in the crystal growth velocities with respect to the bulk, but also an enhancement of the nucleation rate due to heterogeneous nucleation. The results  support the idea of investigating Ge$_2$Sb$_2$Te$_5$/TiTe$_2$ superlattices for applications in neuromorphic devices with improved data retention. The effect on the crystallization kinetics  of the addition of van der Waals interaction to the interatomic potential is also discussed. 
\end{abstract}


\maketitle  

\section{Introduction}
The Ge$_2$Sb$_2$Te$_5$ (GST) compound is the prototypical material for applications in phase change memories (PCMs) in which a binary information is encoded by the crystalline and amorphous phases of GST.\cite{WUTTIG2007,zhangNatMat}
Joule heating through the material or via local heaters leads to either  amorphization via crystal melting (reset operation) or to  recrystallization of the amorphous phase (set operation), while  read out of the memory consists of a measurement of the resistance at low bias that differs by three order of magnitude between the two phases.
Phase change alloys are also among the most promising materials for the realization of artificial neurons and synapses for neuromorphic computing.\cite{sebastian2} In these applications one exploits the different resistivity levels of the material that can be achieved either by  partial crystallization of the amorphous phase in the set process or by varying the size of the amorphous region during reset.

In a recent work,\cite{DingPCH} it was proposed that  a superlattice (SL) geometry made of alternating layers of the phase change material Sb$_2$Te$_3$ and more thermally stable confining layers of TiTe$_2$ exhibits superior properties for neuromorphic computing.
The progressive amorphization or recrystallization of several Sb$_2$Te$_3$ slabs allows for a better control of the different resistance states for neuromorphic applications. Moreover, the read out of the different resistance states in PCMs is typically hampered by a drift with time of the resistivity of the amorphous phase induced by structural relaxations (aging). \cite{ielmini_drift}
This drift can be reduced by nanoconfinement   as it was also shown  in Ref. \cite{DingPCH} for  the Sb$_2$Te$_3$/TiTe$_2$ SL. In the PCM device made of this SL, the TiTe$_2$ slabs act as a thermal and diffusion barrier  that keeps the crystalline form during cycling because of its high melting temperature, while Sb$_2$Te$_3$ undergoes the phase change. \cite{DingPCH} This mechanism of the operation of the memory has been, however, questioned by a recent work \cite{cohen} in which the TiTe$_2$ slabs were shown to disappear upon cycling in the active region of a mushroom cell made of a Sb$_2$Te$_3$/TiTe$_2$  SL.
Anyway, pure Sb$_2$Te$_3$  has a relatively low crystallization temperature that would limit data retention in the memories. Therefore, it has been proposed to 
substitute Sb$_2$Te$_3$ with a phase change material with a higher crystallization temperature such as GST or GeTe.\cite{acharya2024,mazzarello2024}  Sb$_2$Te$_3$ and GeTe are actually the parent compounds of GST that can be seen as a pseudobinary
compound along the GeTe-Sb$_2$Te$_3$ tie-line. We should consider, however, that nanoconfinement could slow down the crystallization kinetics,\cite{KooiReview} 
as it is the case of elemental Sb, for instance, whose amorphous phase crystallizes explosively at 300 K in the bulk, but it is dramatically stabilized in ultrathin films 3-10 nm thick capped by insulating layers. \cite{salinga,DragoniSb} 
In a previous work,\cite{acharya2024} we have shown by  molecular dynamics (MD) simulations that nanoconfinement only slightly reduces the crystallization speed of  GeTe that could then be used for memory applications in the superlattice GeTe/TiTe$_2$ geometry with a foreseen superior data retention with respect to 
Sb$_2$Te$_3$/TiTe$_2$.
It is therefore of interest to investigate whether the flagship GST phase change compound  could be used as well in  GST/TiTe$_2$ SLs.

The structural properties of GST/TiTe$_2$ SLs in the crystalline and partially liquid/amorphous phases have actually been studied in a recent work by atomistic simulations based on Density Functional Theory (DFT), \cite{mazzarello2024}  where it was shown that the interaction between TiTe$_2$ and GST is weak and mostly due to van der Waals (vdW) forces. The stress on the GST film induced by the TiTe$_2$ capping should then be small as well, as opposed to the situations with other capping layers. A capping layer of Al$_2$O$_3$, for instance, was shown to strongly increase the crystallization time for film thickness below 10 nm.\cite{raouxJAP2010} Capping by W
 was  also shown to strongly hinder crystallization in thin  (7 nm) GST films
 at temperatures below about 500 K, while crystallization speed is only slightly affected at higher temperatures.\cite{chenW} In several papers, \cite{orava2012,kolobov2010} it was proposed that  the less facile crystallization at lower temperatures  arises via different possible mechanisms related to the stress induced by the capping layer. 

At higher temperatures, the density change upon crystallization can easily be accommodated  by viscous flow and then the interface with the capping layer has the opposite effect of enhancing heterogeneous crystal nucleation\cite{orava2012}  (see Ref. \cite{KooiReview} for a review and a thorough discussion on crystallization in thin GST films).

On these premises, in this article we report on MD simulations of  the crystallization of GST in a nanoconfined geometry mimicking the superlattice made of alternating layers of  GST and TiTe$_2$, similarly to the Sb$_2$Te$_3$/TiTe$_2$ superlattices of Ref. \cite{DingPCH}. 
 To this aim, we have exploited the neural network (NN) interatomic potential\cite{Behler2007} for GST \cite{npjOmar} that we have recently developed by fitting a large DFT database of energy and forces within the NN  framework implemented in the DeepMD code. \cite{DeePMD4,DeePMD2,DeePMD3}  
 In the present work, the NN potential fitted on the DFT database is supplemented by the semiempirical van der Waals (vdW) correction due to Grimme (D2).\cite{grimmeD2}
Therefore, for the sake of comparison with the SL, we also repeated the simulation of the crystallization in the bulk with vdW corrections, that were not included in our previous work.\cite{npjOmar}

\section{Computational Details} \label{compdet}
Molecular dynamics  simulations  were performed by using the NN interatomic potential  for GST developed in our previous work \cite{npjOmar} by using the DeePMD package.\cite{DeePMD4,DeePMD2,DeePMD3}. The NN was trained on  a DFT database of energies and forces of about 180000 configurations of  small supercells (57-108 atoms) computed by employing the Perdew-Burke-Ernzerhof (PBE) exchange and correlation functional \cite{PBE} and norm conserving pseudopotentials. \cite{GTH1}  The  potential was validated on the structural and dynamical properties of the liquid, amorphous and crystalline phases and it was exploited to study the crystallization kinetics in the bulk. \cite{npjOmar}

At normal conditions, the thermodynamically stable form of GST  is a hexagonal crystal (space group $P{\bar 3}m1$)\cite{petrov,kooi,matsunaga} with nine atoms  in the primitive unit cell arranged along the $c$ direction with a  ABCABC stacking.  Each formula unit forms a lamella separated from the others by a so-called vdW gap, although the interlamella interaction is not just a vdW contact  as discussed in Ref. \cite{Raty2018}. 
Three different models of hexagonal GST have been proposed in literature differing in the distribution of Sb/Ge atoms in the cation sublattices.\cite{petrov,kooi,matsunaga}
Here, we considered the Kooi stacking \cite{kooi} with Sb atoms occupying the cation planes close to the vdW gap and without disorder in the cation sublattice.
It was shown that  GST features phonon  instabilities in the Kooi stacking when the DFT-PBE scheme is applied.\cite{campi} 
These instabilities at the PBE level are removed \cite{campi} by including the semiempirical vdW correction due to Grimme (D2). \cite{grimmeD2}
In the crystallization process, the amorphous (supercooled liquid) phase transforms into the cubic phase which is instead stable at the PBE level.  Therefore, in our previous work on the crystallization of bulk GST,\cite{npjOmar}  vdW interactions were not included.
In the present work, we will instead start from the hexagonal phase of GST to model the superlattice geometry.
Moreover, vdW interactions are  expected to control the interplanar distance between the GST slab and the TiTe$_2$ slabs. Therefore, in the present simulations we have added the vdW-D2 interactions.
In the next section, we will briefly summarize results on the crystallization kinetics in the bulk by including vdW interactions, before moving to the discussion of the crystallization of GST in nanoconfined geometry.

\begin{figure}[t]
\centering
 {\includegraphics[height=10cm]{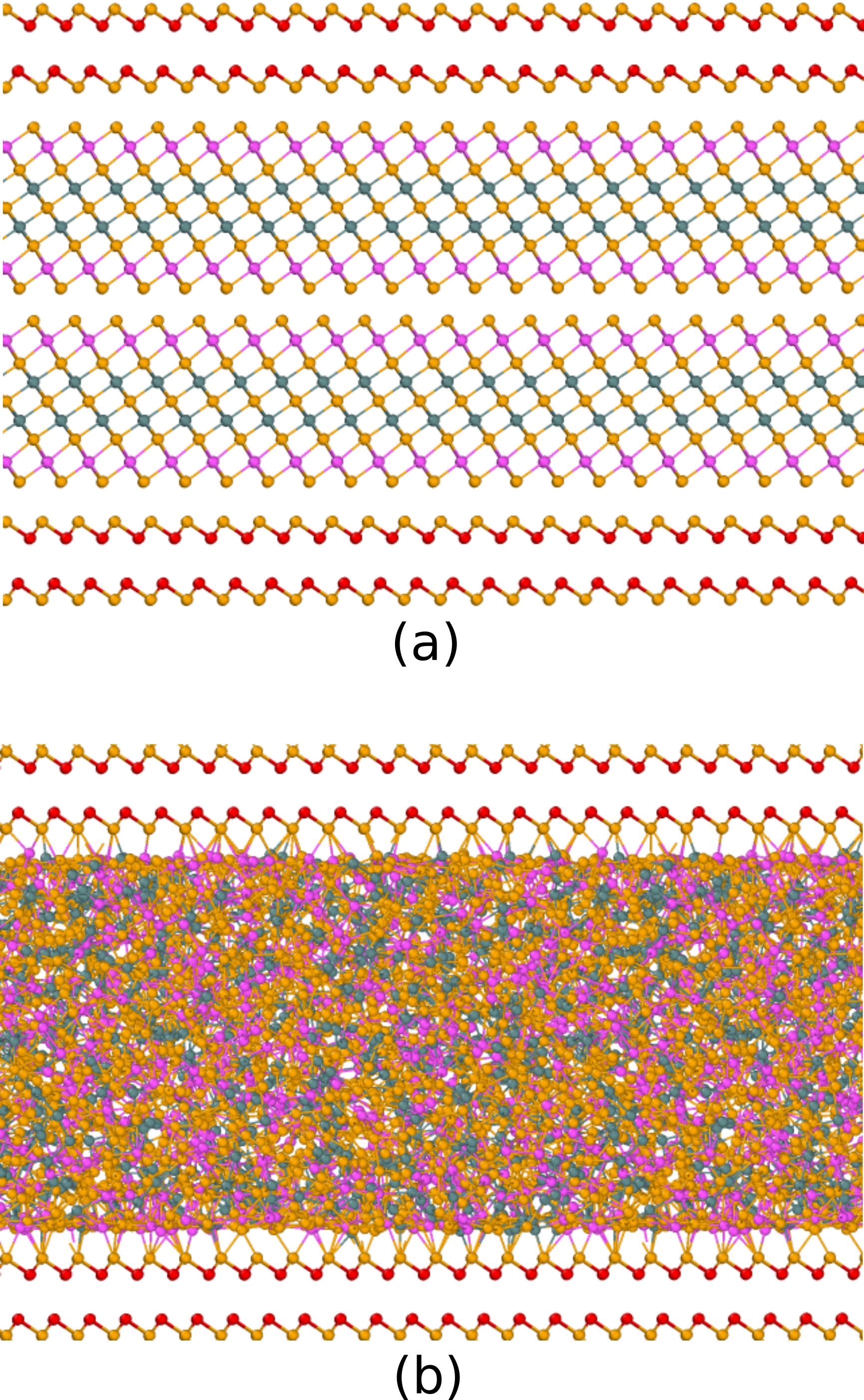}}
\\
\caption{(a) Crystalline and (b) amorphous phase of the slab made of  two quintuple layers of GST encapsulated by capping layers. The capping layer is  made by a frozen bilayer of crystalline GeTe at the lattice constant of  TiTe$_2$, aiming at mimicking the confining slabs of TiTe$_2$ in GST/TiTe$_2$ SLs. Color code for atomic spheres: Ge (gray), Te (orange), Sb (pink), Ge atoms in the capping layers (red). The Ovito\cite{ovito} tool was used for the visualization and the generation of all atomic snapshots of this article.}
\label{fig:GSTTiTe2model}
\end{figure}

We have simulated the effect of confinement on the crystallization of GST by considering a slab made of two
quintuple layers  of  GST (18 atomic planes with a thickness of  about 3 nm), encapsulated by capping layers aiming at mimicking the confining slabs of TiTe$_2$ in GST/TiTe$_2$ SLs. 
As we did in a previous work on nanoconfined GeTe,\cite{acharya2024}
the capping layer mimicking TiTe$_2$  on each side is made by a frozen bilayer of crystalline GeTe itself constrained at the lattice parameter of TiTe$_2$ as shown in Fig. \ref{fig:GSTTiTe2model}a. In fact,  TiTe$_2$ is a layered hexagonal crystal (space group P$\bar{3}$m1) made of trilayer Te-Ti-Te blocks  stacked along the $c$ axis and separated by vdW gaps.\cite{greenaway1965preparation}  The geometry of the hexagonal Te layers is the same in TiTe$_2$ and GST albeit with  different lattice parameters.
The $a$ and $c$ lattice parameters of hexagonal GST computed with and without vdW corrections are compared in Table \ref{lattice-parameters} with experimental data.\cite{matsunaga}  The experimental in-plane lattice parameter of TiTe$_2$ is instead a=3.7795 \AA.  \cite{greenaway1965preparation} A good commensuration between a trilayer of TiTe$_2$ and GST in the hexagonal $xy$ plane is obtained by considering multiples of the orthorhombic supercells with edge $a$ and $\sqrt{3}a$, namely  16$\times$8 orthorhombic cells of TiTe$_2$ and 14$\times$7 orthorhombic cells of  GST. The misfit is only 2 $\%$ along both $x$ and $y$.  We finally set the in-plane lattice parameters of the supercell to those of  GST which means that the bilayers mimicking TiTe$_2$ are slightly strained by the amount given above.  The model thus contains 14$\times$7$\times$2 = 196 atoms per atomic layer of GST  and 16$\times$8$\times$4 = 512 atoms in each of the bilayers mimicking TiTe$_2$, for a total amount of 4552 atoms of which 3528 are mobile. Periodic boundary conditions are applied along the three cartesian axis (see Fig. \ref{fig:GSTTiTe2model}). The TiTe$_2$--like bilayers are oriented in such a way to expose the Te layer to the GST slabs on both sides as it would occur for a TiTe$_2$ slab. The distance between the outermost Te plane of the capping layer and of hexagonal GST  is fixed to the value of 3.55
\AA, obtained from geometry optimization at the PBE-D3 level of a GST/TiTe$_2$ SL in Ref.\cite{mazzarello2024}.

\begin{table}
  \begin{tabular*}{0.48\textwidth}{@{\extracolsep{\fill}}cccc}
    \hline
 &NN+vdW &NN &Expt\\
\hline
$a$ (\AA) &4.235 &4.30 & 4.2247\\
$c$ (\AA) &17.15 &17.64  &17.2391\\
$\rho$ (atom/\AA$^{3}$) &0.03379 & 0.03182 & 0.03378\\
    \hline
    
  \end{tabular*}

\caption{Theoretical equilibrium lattice parameters $a$ and $c$ (\AA) and equilibrium density $\rho$ (atom/\AA$^{3}$) of hexagonal GST computed with (NN+vdW) and without (NN) van der Waals corrections, compared with experimental data from Ref. \protect\cite{matsunaga}. Notice that there was a misprint in the NN equilibrium volume in Table 3 of Ref. \cite{npjOmar}. The correct value was actually equal to the DFT result, as it can be inferred from the Supplementary Figure 8 of Ref. \cite{npjOmar}.}
\label{lattice-parameters}
\end{table}

Since TiTe$_2$ has a much higher melting temperature than GST, we mimicked the confinement by TiTe$_2$ by freezing the atoms of the crystalline GeTe-like capping bilayers during the thermal cycle. 

MD simulations were performed with the DeePMD code by using the Lammps code as MD driver,\cite{LAMMPS} a time step of 2 fs, and  a Nos\'e-Hoover thermostat.\cite{noseart,hoover}  

The GST slab was amorphized by first equilibrating the system at 1500 K for 200 ps and then at 1000 K for 100 ps. The liquid-like slab was then quenched to 300 K in 150 ps. 
Structural properties of the resulting amorphous slab were computed over 70 ps simulation at 300 K. 
The amorphous model was then heated at different target temperatures to study the crystallization process in simulations about 1-3 ns long at each temperature at constant volume. 
To identify the crystalline nuclei we used the local order parameter  $Q_{4}^{dot}$,\textsuperscript{\cite{Q6,q4}} that we considered in our previous work on the crystallization of bulk GST.\cite{npjOmar}

\section{Results and Discussion}   \label{results}  

\subsection{Structural properties of the bulk}
The structure of bulk amorphous GST (a-GST) is discussed in previous DFT works \cite{caravati225,akola2007} and in our previous work on the development of the NN potential.\cite{npjOmar}
Te atoms are mostly three-fold
coordinated in a pyramidal geometry, Sb atoms are both three-fold coordinated in a pyramidal geometry (three bonding angles of 90$^{\circ}$) and four- or five-fold
coordinated in a defective octahedral environment (octahedral
bonding angles but coordination lower than six), 
Ge atoms are mostly in pyramidal or defective octahedral geometry
with a minority fraction in tetrahedral geometries. 

\begin{figure*}[t]
\centering
\includegraphics[height=9.0cm]{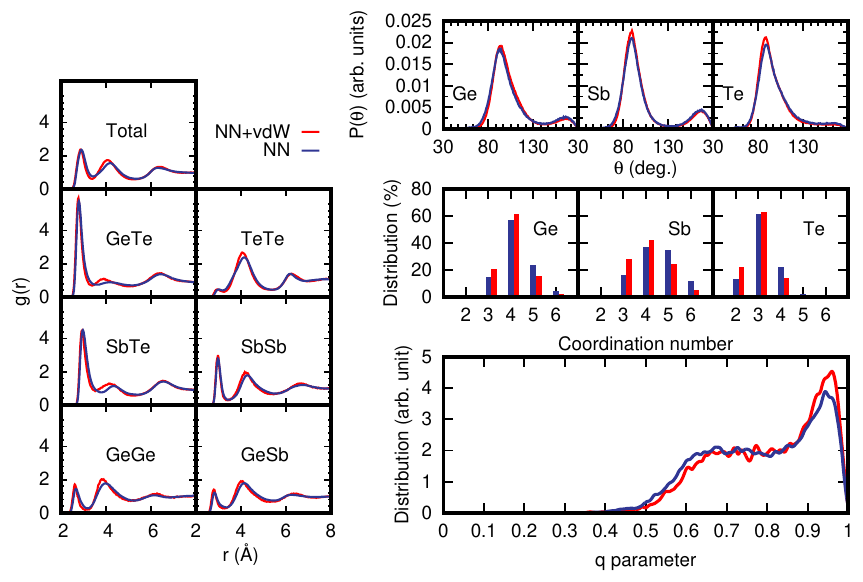}
\caption{Structural properties at 300 K of models of amorphous GST 
generated with (red lines) and without (blue lines from Ref. \cite{npjOmar})  vdW corrections, both at the experimental density of the amorphous phase of 0.0309 atom/\AA$^3$ (bulk-LD). (a) Partial pair correlation functions.  (b) Distribution of coordination numbers resolved per chemical species. (c) Bond angle distribution function resolved per central atomic species. The data are normalized to the number of triplets in each model. (d) Distribution of the q order parameter for tetrahedricity of the fourfold coordinated Ge atoms.}
\label{fig:structurebulk}
\end{figure*}

A comparison of the structural properties of a-GST with and without vdW interaction at the experimental density of the amorphous phase (0.0309 atom/\AA$^3$)\cite{njoroge2002density} is shown in Fig. \ref{fig:structurebulk}. We name this model a low density configuration (bulk-LD) to distinguish it from other models at higher density that we will discuss later on. 
The data with no vdW interaction are taken from our previous work. \cite{npjOmar}, while the data with vdW interaction are obtained from the simulation of a 3996-atom cubic model generated by quenching from 1000 K to 300 K in 100 ps.
Pair correlation functions, angle distribution functions and the distribution of coordination numbers are shown in Fig. \ref{fig:structurebulk}, while the average partial coordination numbers are compared in Table S1 in the ESI\dag. 
The coordination numbers are obtained by integrating the partial pair correlation functions up to the bonding cutoff of 3.2 {\AA } (Ge-Ge, Ge-Sb, Ge-Te, Sb-Te and Te-Te) and  3.4 {\AA } (Sb-Te). A quantitative measure of the fraction of tetrahedral environments can be obtained from the local order parameter $q$ introduced in 
Ref.~\cite{errington}. It is defined as $q=1-\frac{3}{8}\sum_{\rm i > k} (\frac{1}{3} + \cos \theta_{ \rm ijk})^2$,   where the sum runs over the pairs of atoms bonded to a central atom $j$ and 
forming a bonding angle $\theta_{\rm ijk}$.The order parameter evaluates to $q$=1 for the ideal tetrahedral geometry, to $q$=0 for the 6-fold coordinated octahedral site,
to $q$=5/8 for a 4-fold coordinated defective octahedral site, and $q$=7/8 for a pyramidal geometry. 
The distribution of the local order parameter $q$ for tetrahedricity  for four-coordinated Ge atoms is also reported in Fig. \ref{fig:structurebulk}d. The bimodal shape corresponds to tetrahedral and defective octahedral geometries. We quantified the fraction of Ge atoms in a tetrahedral environment by integrating the $q$ parameter between 0.8 and 1 as discussed in  previous works.\cite{spreafico} The fraction of Ge atoms in a tetrahedral geometry is almost equal in simulations with (33\%) and without (30\%) vdW corrections.

\begin{figure}[h]
\includegraphics[width=0.5\textwidth,keepaspectratio]{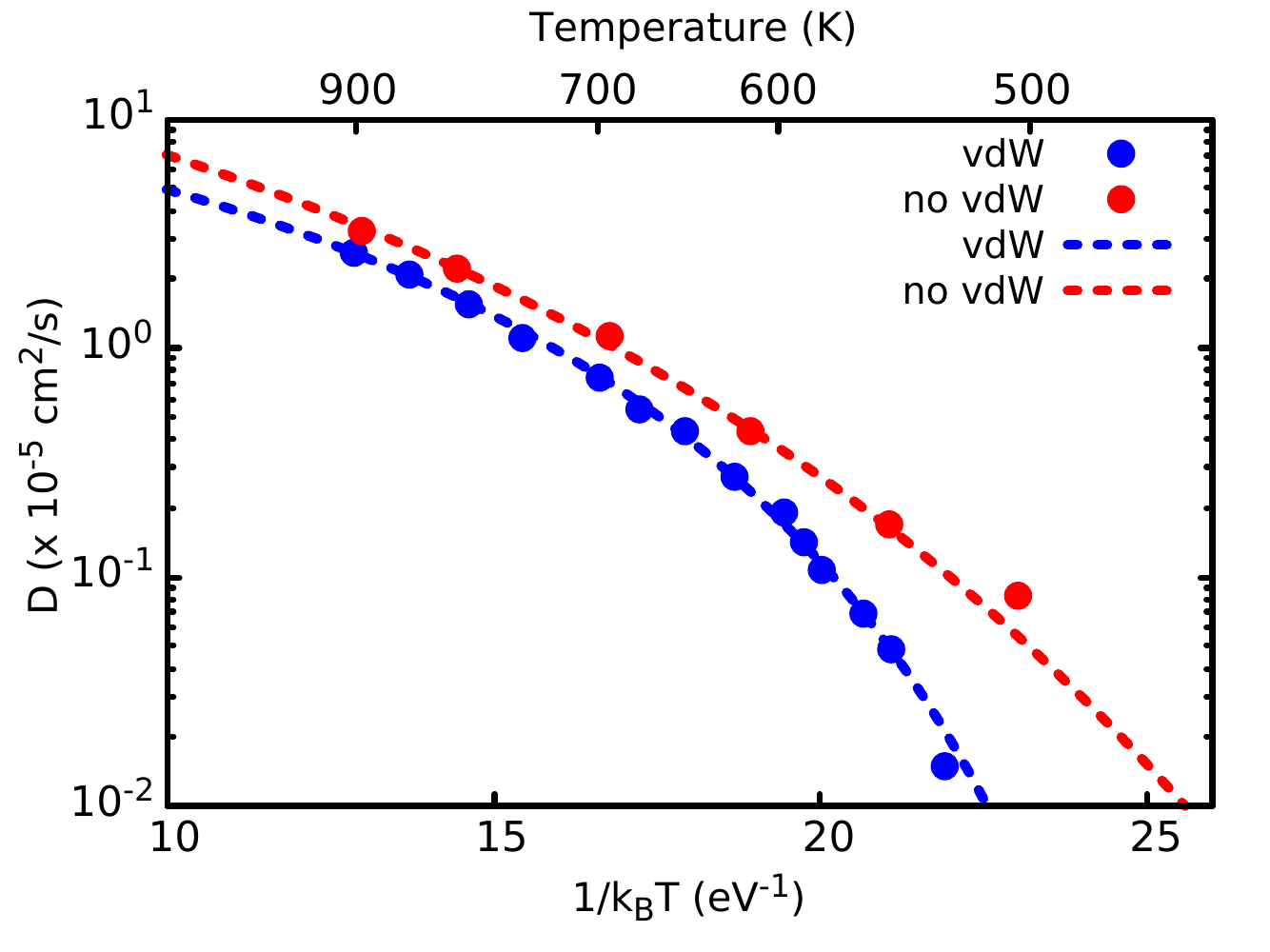}
\includegraphics[width=0.5\textwidth,keepaspectratio]{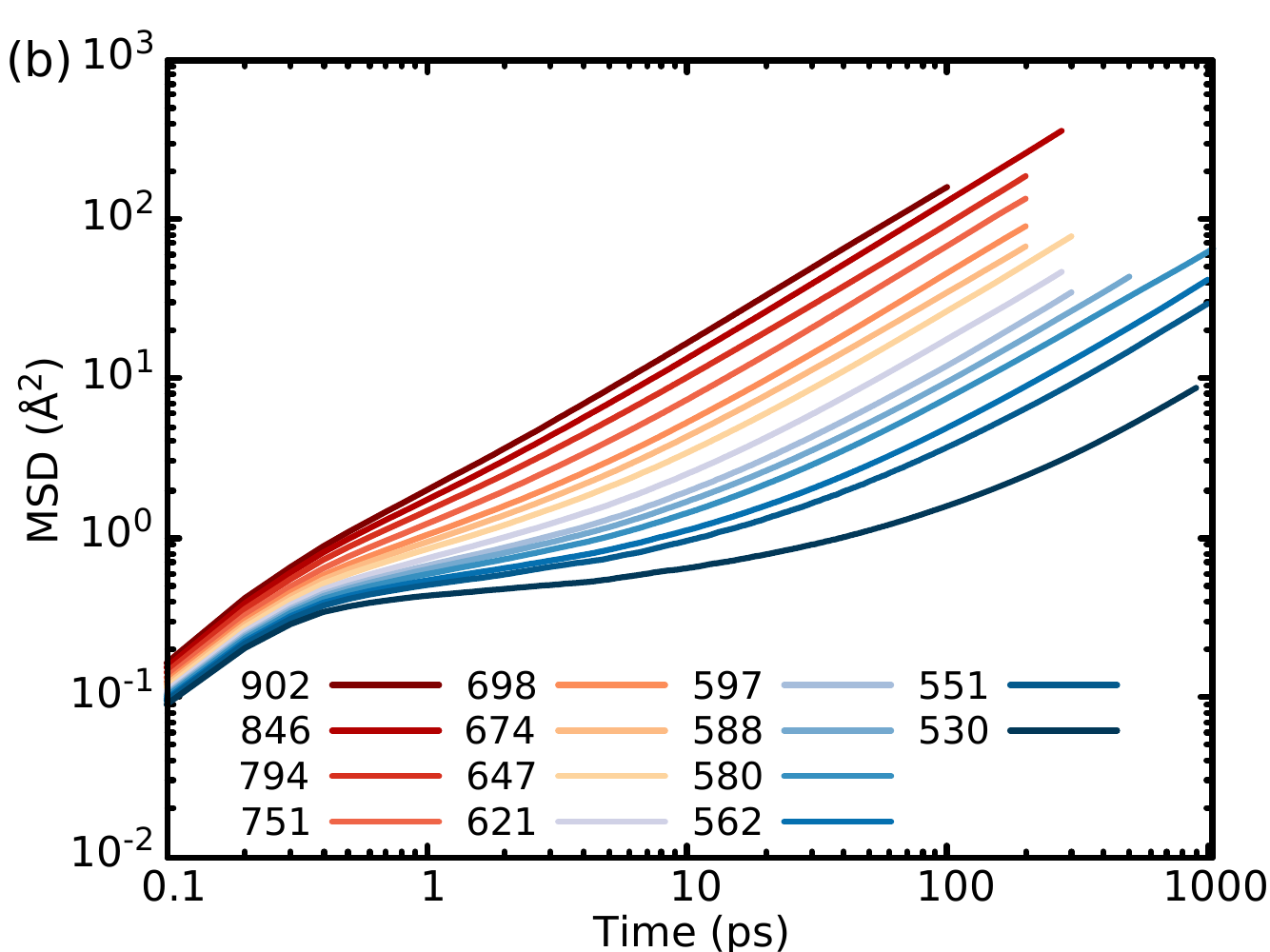} 
\caption{(a) Diffusion coefficient $D$ as a function of temperature from bulk NN simulations with  and without  vdW interactions at the experimental density of the amorphous phase (0.0309 atom/\AA$^3$). The data (points) are fitted by the Cohen-Grest  (CG) formula \cite{CG}  as $\log_{\rm 10}(D(T))=A-{2B}/(T-T_{\rm 0} + [(T-T_{\rm 0})^2 + 4CT ]^{1/2})$, which yields  $A$=-2.45, $B$=602 K, $C$=17.3 K and $T_{\rm 0}$=330.6 K without vdW (see Ref. \protect\cite{npjOmar}) and  $A$=-3.76, $B$=433 K, $C$=0.76 K and $T_{\rm 0}$=384.5 K with vdW (dashed lines). (b) Mean square displacement (MSD) of bulk GST as a function of time at different temperatures (K) from NVE simulations (NN+vdW) in the supercooled liquid.}
\label{diffusion}
\end{figure}

 The addition of vdW interactions leads to a slightly better defined first minimum (first coordination shell) of the pair correction functions both in the amorphous and in the liquid phases (not shown here), as it was already observed for GeTe. \cite{micoulaut,acharyaNPs2024} On the other hand,  vdW corrections have a strong impact on the atomic mobility, as already found in previous works on GeTe as well.\cite{weber} Indeed, the diffusion coefficient $D$ in the supercooled liquid is  around a factor three lower with the addition vdW corrections. The diffusion coefficient with and without vdW interactions are compared in Fig.  \ref{diffusion}a. 
The diffusion coefficient was obtained from the mean square displacement (MSD) and the Einstein relation MSD=6$Dt$ from equilibrated trajectories at constant energy (NVE simulations). The MSD as a function of time at different temperatures is shown in Fig. \ref{diffusion}b.

\subsection{Structural properties of the superlattice}

The a-GST slab encapsulated by the frozen capping bilayers mimicking TiTe$_2$ (superlattice configuration) is shown in Fig. \ref{fig:GSTTiTe2model}b.

We observed a small expansion of the amorphous slab which leads to a density of about 0.0329 atom/\AA$^3$ (estimated in spheres of 10 {\AA } radius in the inner part of the slab) to be compared with  the initial density of 0.0338 atom/\AA$^3$ of the slab in the crystalline phase. 
We remark that the experimental density of the amorphous phase is 
 0.0309 atom/\AA$^{3}$.\cite{njoroge2002density}

\begin{figure*}[ht]
\centering
     {\includegraphics[height=9.0cm]{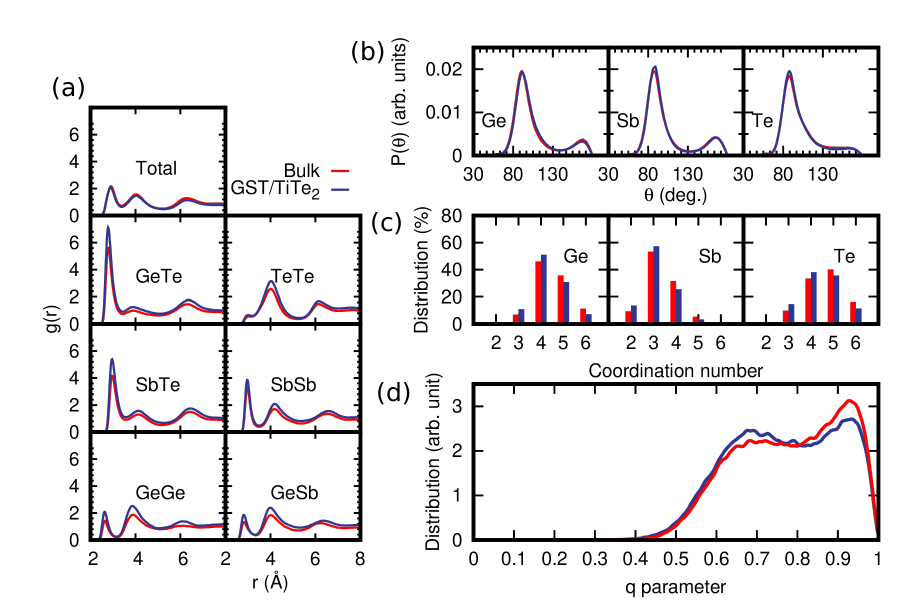}}
\caption{Structural properties at 300 K of the slab of amorphous GST  confined by the TiTe$_2$-like capping layers in SL (blue lines, SL, see Fig.\ref{fig:GSTTiTe2model}b) compared to those of a bulk model of amorphous GST at the density of the hexagonal phase generated by quenching from the melt (bulk-HD, red lines). (a) Partial pair correlation functions.  (b) Distribution of coordination numbers resolved per chemical species. (c) Bond angle distribution function resolved per central atomic species. The data are normalized to the number of triplets in each model. (d) Distribution of the q order parameter for tetrahedricity of the fourfold coordinated Ge atoms.}
\label{fig:structure}
\end{figure*}

 The structural properties of the a-GST slab are compared
 in Fig. \ref{fig:structure}
 to those of a bulk model quenched from the melt at the  density fixed to the theoretical density
of hexagonal GST (with vdW, see Table 1). 
This high density bulk model (bulk-HD) was generated by quenching from 1000 K to 300 K in 150 ps a 3528-atom orthorombic cell with lattice parameters $a$= 59.29 {\AA }, $b$= 51.34 {\AA }, and $c$= 34.30 {\AA }.
Partial pair correlation functions, bond angle distribution functions, 
distribution of the coordination numbers, and the distribution of order parameter $q$ for tetrahedricity are shown in Fig. \ref{fig:structure}a-d. The bonding cutoffs are the same as those used for the bulk in the previous section. The average partial coordination numbers of the slab and the bulk are compared in Table \ref{tab:Tablecoor}.

\begin{table}[h]
\small
  \caption{Average coordination number for different pairs of atoms computed from the partial pair correlation functions for the amorphous slab confined by the capping layers (GST/TiTe$_2$-like SL), compared with the data of a bulk amorphous model at the density
of the hexagonal phase of GST (bulk-HD, see text).}
  \label{tab:Tablecoor}
  \begin{tabular*}{0.48\textwidth}{@{\extracolsep{\fill}}rcccc}
    \hline
    & & &bulk-HD &GST/TiTe$_2$-like SL\\
    \hline
&Ge &with Ge &0.36 &0.38\\
& &with Sb &0.39 &0.38\\
& &with Te &3.76 &3.58\\
& &Total &4.52 &4.34\\
&Sb &with Ge &0.39 &0.38\\
& &with Sb &0.74 &0.65\\
& &with Te &3.51 &3.41\\
& &Total &4.63 &4.44\\
&Te &with Ge &1.50 &1.43\\
& &with Sb &1.40 &1.36\\
& &with Te &0.43 &0.39\\
& &Total &3.33 &3.18\\
    \hline
  \end{tabular*}
\end{table}

The coordination numbers are lower in the superlattice than in the bulk in part because of the slightly lower density (0.0329  vs 0.0338 atom/\AA$^3$) and because of the presence of the two surfaces.
We observed a slight enrichment of Te 
(66 $\%$ instead of 56 $\%$) at the two surfaces of the amorphous slab facing the Te planes of the capping layers. 
The resulting fraction of Ge atoms in tetrahedral geometry (with respect to the total number of Ge atoms) is 24.5 $\%$ in the slab and 20.1 $\%$ in the amorphous model of the bulk at the density of the hexagonal crystal.

\subsection{Crystallization kinetics in bulk Ge$_2$Sb$_2$Te$_5$}
\begin{figure}[h!]
 \includegraphics[width=0.5\textwidth,keepaspectratio]{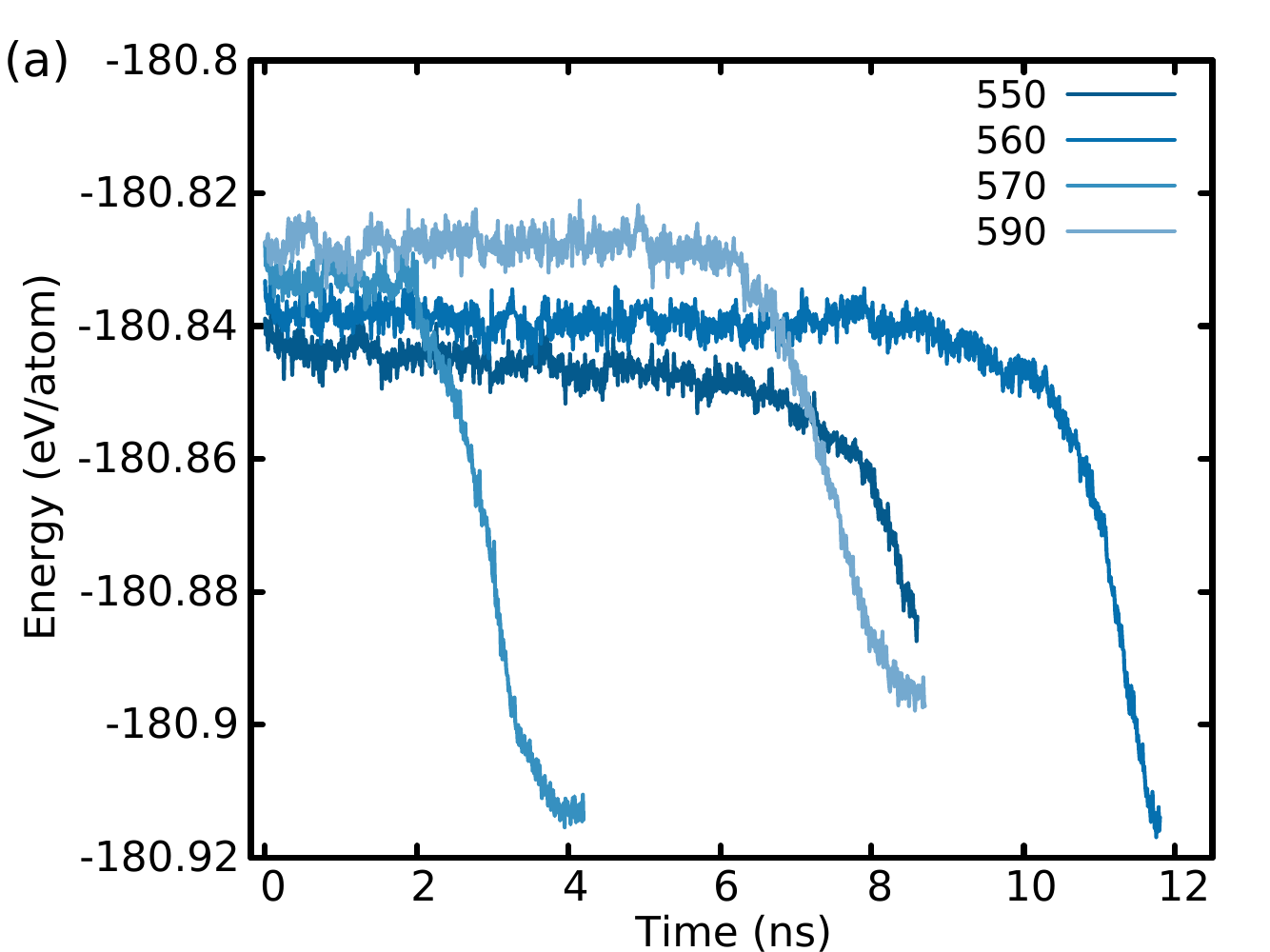} 
 \includegraphics[width=0.5\textwidth,keepaspectratio]{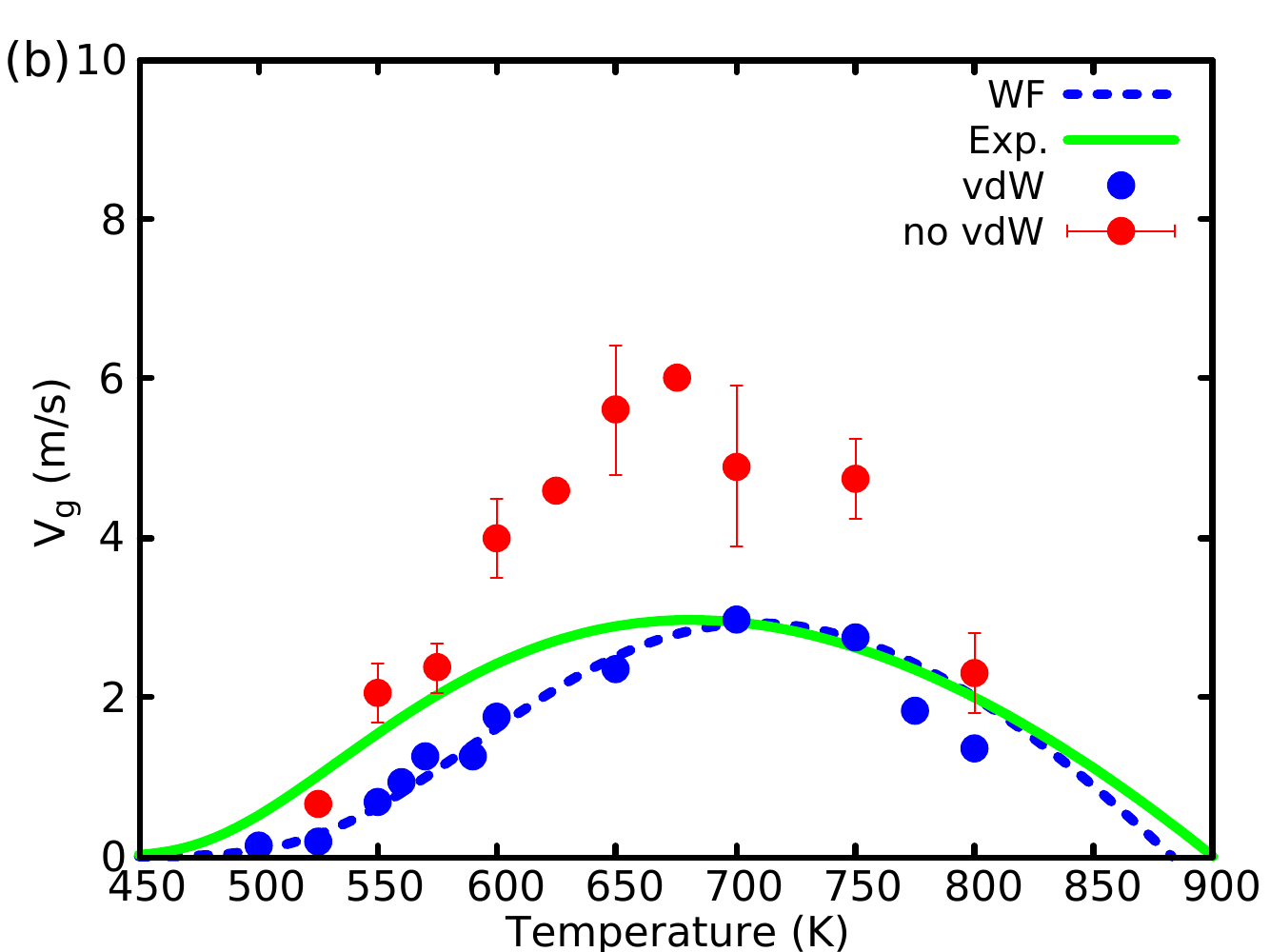}
\caption{(a) Potential energy as a function of time in NN+vdW simulations of the homogeneous crystallization of the supercooled liquid phase at different temperatures and at the experimental density of a-GST (bulk-LD). (b) Crystal growth velocities from simulations with (blue dots) and without (red dots from Ref. \protect\cite{npjOmar}) vdW corrections. 
The error bars when present refer to data extracted from more than two nuclei. 
Experimental data from ultrafast differential scanning calorimetry (green curve)\protect\cite{orava}  and the fitting of the NN+vdW data with the WF formula 1  (blue dashed line)  are also shown.}
\label{nucleationBulk}
\end{figure}

The kinetic of crystallization in the bulk was analyzed in our previous work \cite{npjOmar} from simulations without vdW corrections.
To study crystallization with vdW interaction, we generated a 3996-atom cubic model at the experimental density of the amorphous phase (see above).
The model was first equilibrated at 1200 K for 40 ps, then quenched to 900 K in 40 ps and further equilibrated for 60 ps. The system was then brought to the target temperature in 160 ps to study nucleation and growth.  For temperatures where nucleation was not observed after a few nanoseconds, the crystal growth velocities were estimated by heating (cooling) at the target temperature a configuration with an overcritical nucleus generated at a lower (higher) temperature. 
The reduction of the self-diffusion coefficient upon addition of the vdW interactions, mentioned above, leads to an increase of the nucleation time and to a decrease of the crystal growth velocity.
The potential energy as a function of time for NN+vdW simulations at different temperatures, shown in Fig. \ref{nucleationBulk}a, reveals the onset of the crystallization with a nucleation time that increases with temperature. Overcritical nucleus/nuclei form on a time scale of 4-12 ns in the temperature range 550–590 K. Nucleation was not observed at and above 600 K in simulations lasting over 12 ns. 

The crystal growth velocity was computed as $v_{\rm g}$=${dR(t)/dt}$  by assuming a spherical  overcritical nuclei with  radius $R$ given by $R\left(t\right)$=$\left({3N(t)}/({4\pi\rho_{cubic}})\right)^{\frac{1}{3}}$, where $N$ is the number of atoms in the nucleus, $\rho_{cubic}$ is the theoretical density of the cubic crystal (0.0331 atoms/\AA$^3$). The evolution in time of the radius of overcritical nuclei at different temperatures are shown in Fig. S1 in ESI\dag.
Crystal growth velocities with and without vdW interactions for bulk a-GST at the experimental density (bulk-LD) are compared in Fig. \ref{nucleationBulk}b. The results with vdW corrections are in an overall better agreement with experimental data from ultrafast different scanning calorimetry of Ref. \cite{orava}. The reduction of $v_{\rm g}$ due to the vdW correction mirrors the reduction of the diffusion coefficient reported in Fig.
\ref{diffusion}.

The theoretical $v_{\rm g}$  was fitted by the Wilson-Frenkel (WF) formula as we did in our previous work \cite{npjOmar} for the simulations without vdW corrections. Namely, we used the WF formula
$v_g(T)= 6 D(T)d/\lambda^2 (1 - e^{(-\Delta \mu(T) /k_BT)}) e^{-\Delta S(T)/k_B}$, where $d$ is a geometric factor that we will define later, 
$\lambda$ is a typical jump distance which is used as fitting parameter, and $D(T)$ is the temperature dependent diffusion coefficient extracted from the simulations. 
$\Delta S$ is the (positive) entropy difference between the liquid  and the crystal which is computed from the specific heat (with vdW), as discussed in our previous work \cite{npjOmar} to which we refer to for further details. The factor $e^{-\Delta S/k_B}$ in the WF formula, which  was first introduced by Jackson, \cite{jackson} turned out to be necessary to reproduce the theoretical $v_g(T)$ without vdW corrections in our previous work.\cite{npjOmar} 
$\Delta \mu(T)$ is the difference in free energy between the liquid and the crystal given by the Thompson-Spaepen approximation $\Delta\mu(T)=\frac{\Delta H_{m}(T_m-T)}{T_m}\frac{2T}{(T_m+T)}$,\cite{thompson1979approximation} where $\Delta H_{m}$ is the enthalpy jump at the melting temperature ${T_m}$. We set $\Delta H_{m}$ = 166 meV/atom and $T_m$ = 940 K as estimated in our previous work from NN+vdW simulations.\cite{npjOmar} Due to the sensitivity of the results on the choice of $T_m$, we have also further checked for possible finite size effects in the estimate of $T_m$  from the phase coexistence method in the slab geometry used in our previous work.\cite{npjOmar} By increasing the $c$ axis of the slab model from the value of 11 nm of Ref. \cite{npjOmar} to 30 nm, the resulting $T_m$ of 940 K does not change. At the highest temperatures,  $\Delta \mu(T)$  becomes very small and therefore
the corrections due to the surface energy of the crystalline nucleus become important. 
In fact,  the change in free energy due to the addition of an atom to a spherical nucleus is given by -$\Delta \mu(T) + 2\sigma/(\rho_{cubic} R)$ where $\sigma$ is the crystal-liquid interfacial energy. Indeed, at the highest temperatures above 700 K,
$R(t)$ changes slope with time   due to the interfacial term, as shown in Fig. S1 in the ESI\dag.
Since $v_{\rm g}$  was computed from $R$ in the range 15-20 {\AA } at all temperatures, in the lack of a reliable estimate of $\sigma$, we
considered the term $2\sigma/(\rho_{cubic} R)= \Delta \mu_S$ as a constant offset to  -$\Delta \mu(T)$ which was used as an additional fitting parameter in the WF formula.

\begin{figure*}[]
\centering
 {\includegraphics[height=6cm]{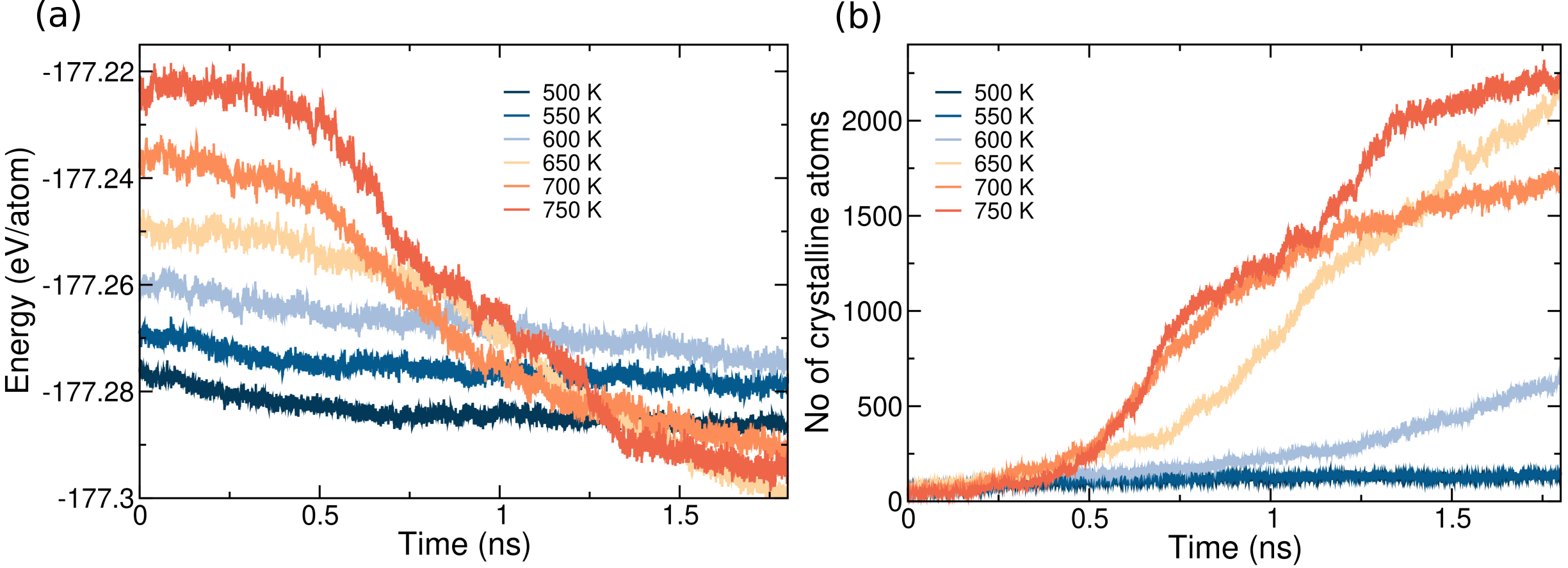}}
\caption{Crystallization of amorphous GST capped by bilayers mimicking confinement by TiTe$_2$ in GST/TiTe$_2$-like SL
at different temperatures (SL-HD).  (a) Potential energy and (b) number of crystalline atoms as a function of time. At 550 K crystal nucleation occurs after 2.2 ns, i.e. on a longer timescale not shown in the figure.}
\label{fig:internalenergy}
\end{figure*}

Coming now to the geometric factor,
for a spherical nucleus $d$=$2/3 (vol_{site}3 / (4 \pi))^{1/3} $, where $vol_{site}$ is the volume associated with an adsorption site on the crystalline nucleus.\cite{raoux-ielmini} If we take $vol_{site}=4 \pi /3 (\lambda/2)^3$, the WF formula reads $v_g =4D/\lambda (1- e^{-\Delta \mu/k_B T})e^{-\Delta S(T)/k_B}$. As we did in our previous work,\cite{npjOmar} we here use the more general and complete formula:

\begin{align}
&v_g = u_{kin} (1 - e^{(-\Delta \mu +\Delta \mu_S)/k_BT})e^{-\Delta S(T)/k_B}\ \label{WF}&\\
&with\ u_{kin} = 8 (vol_{site}3 / (4 \pi))^{1/3} D/\lambda^2
\end{align}

\noindent where $vol_{site}^{1/3}$
is about half the lattice parameter of the cubic cell, i.e 3 \AA. 
The resulting fit is shown in Fig. \ref{nucleationBulk}b
with  $\lambda$= 2.04 {\AA }, which is a reasonable jump distance,
and $\Delta \mu_S$=0.0098 eV which corresponds to $\sigma$= 0.052 J/${m^2}$ for $R$= 20 {\AA }. This value is similar to those used previously in numerical simulations (0.060-0.075 J/m$^2$).\cite{oravapriming,burr2012}

\subsection{Crystallization kinetics in the superlattice}
Turning now to SL geometry, the crystallization kinetics  was studied by
  heating the system at six different target temperatures of 500, 550, 600, 650, 700, and 750 K in constant volume simulations, 1-3 ns long each. 
Albeit there is a large mismatch between the lattice parameters of GST and of the capping layers,  crystal nucleation always starts at the surfaces 
as it was observed in  our previous works on nanoconfined GeTe \cite{acharya2024} and  GeTe nanoparticles.\cite{acharyaNPs2024}

Within the classical nucleation theory of heterogeneous nucleation, the crystallites preferentially form at the surface when $\sigma > \sigma_c -\sigma_a$ where  $\sigma_c$ and $\sigma_a$ are surface energy of the crystal and amorphous (supercooled liquid) phases.\cite{CNT,CNT2} 
As discussed in Secs. 2 and 3.2, the density of GST in the SL is close to the theoretical equilibrium density of the hexagonal phase. We call SL-HD the high density model discussed so far to distinguish it from another model at lower density that we will introduce later on. 

The onset of crystallization is visible
from the evolution in time of the  potential energy 
and of the number of crystalline atoms shown in Fig. \ref{fig:internalenergy}.
The nucleation time  of about 0.5 ns at 650-750 K is much shorter than in the bulk (see Fig. \ref{nucleationBulk}) due to heterogeneous nucleation. At 550 K just one nucleus forms at one surface on a longer time scale  of 2.2 ns (not shown in Fig. \ref{fig:internalenergy}),  while at all other temperatures we see the formation of an overcritical nucleus at both surfaces. At 700 K, up to four overcritical nuclei form.
At the highest temperature of 750 K, the crystallization of the slab in nearly complete in 2 ns.

\begin{figure*}[t]
\centering
{\includegraphics[height=12cm]{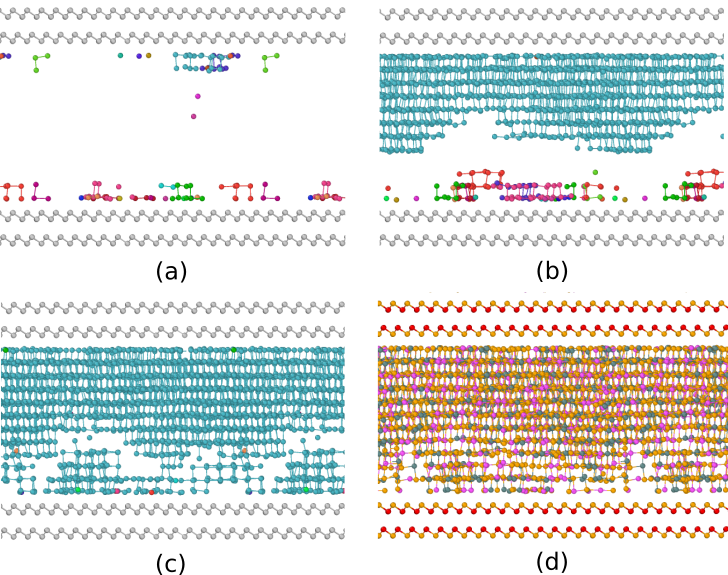}}
\\
\caption{Simulation of the crystallization of a 3528-atom slab of amorphous GST capped by bilayers mimicking confinement by TiTe$_2$ in GST/TiTe$_2$-like SL (SL-HD). Snapshots at different times at 750 K are shown for (a) 0.5 ns (b) 1 ns, and  (c) 1.5 ns. Crystallization starts at the surfaces of the amorphous slab, albeit the capping layers do not act as nucleation centers. Only crystalline atoms, identified by the $Q_4^{dot}$ order parameter (see Sec. 2), are shown. Different crystalline nuclei  have different colors. (d) Final configuration after 2 ns. The color code for panel (d) is the same of Fig. 1.}
\label{fig:Cryst750}
\end{figure*}

Snapshots of the crystallization process at 750 K are shown in Fig. \ref{fig:Cryst750}; similar snapshots for the simulations at 650 and 700 K are given in Figs. S2-S3 in the ESI\dag.  Note that aside the overcritical nuclei, several small undercritical nuclei form and disappear at both surfaces. The crystallites formed at the surface mostly expose the (001) plane of the cubic phase.  For the single temperature of 700 K, we repeated the simulations of crystallization for two other independent amorphized models. In just one case, we also see the formation of a nucleus exposing the (111) plane at the surface.

\begin{figure*}[t]
\centering
{\includegraphics[height=12cm]{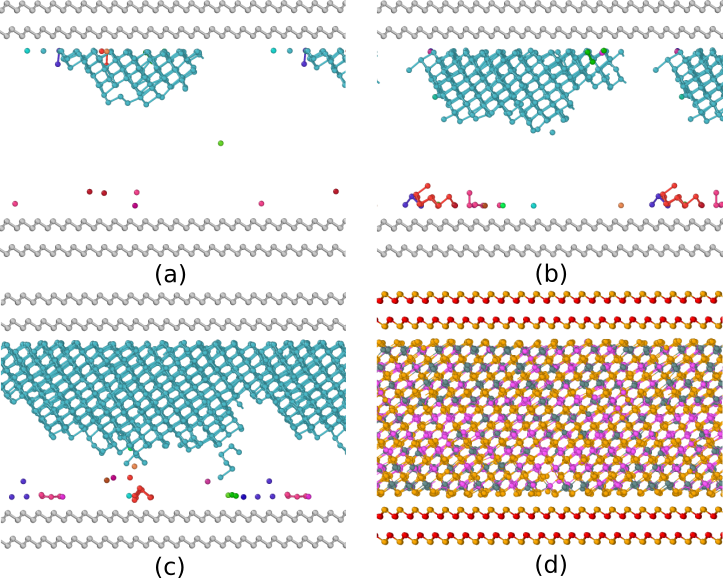}}
\\
\caption{Simulation of the crystallization of a 3528-atom slab of amorphous GST capped by bilayers mimicking confinement by TiTe$_2$ in GST/TiTe$_2$ superlattices at a lower density of the GST slab (SL-LD', see text). Snapshots at different times at 750 K are shown for (a) 0.5 ns (b) 0.75 ns, and  (c) 1 ns. Crystallization starts at the surfaces of the amorphous slab, albeit the capping layers do not act as nucleation centers. Only crystalline atoms, identified by the $Q_4^{dot}$ order parameter (see Sec. 2), are shown. Different crystalline nuclei  have different colors. (d) Final configuration after 1.2 ns. The color code is the same of Fig. 1.}
\label{fig:Cryst750LD}
\end{figure*}

To further analyze the orientation of the crystallites,  we  repeated a simulation for a SL model 
generated by amorphizing a hexagonal phase with a slightly lower density corresponding to the equilibrium lattice parameters obtained without the vdW correction (see Table \ref{lattice-parameters}). We name this model
at lower density SL-LD', where the prime is meant to distinguish this low density from the bulk  model at the experimental density of the amorphous phase
(bulk-LD).
Snapshots of the crystallization process of the SL-LD' model  at 
750 K are shown in Fig. \ref{fig:Cryst750LD}.

The evolution in time of the 
potential energy and of the number of crystalline atoms for this second set of simulations of the
SL-LD' model are shown in Fig. S4 in the ESI\dag .
At 750 and 600 K, we see the formation of an overcritical nucleus just at one surface, while at 700 and 650 K a single overcritical nucleus forms at both surfaces.
During the crystallization process, we first observed an enrichment in Te in the outermost layers of the amorphous slab
which leads to the nucleation at the surface of cubic crystallites all exposing the (111) plane. This occurs at both surfaces of the amorphous slab. 
At 700 K, we repeated the simulations for other two independent models with very similar results.  This preferential orientation of the crystallites  is in agreement with experimental findings on surface nucleation in thick GST films,\cite{kooi2004} albeit at temperatures lower than those simulated here.

We speculate that the different orientation of the crystallites formed in the two sets of simulations at slightly different densities arise from the requirement of minimizing the surface energy of the crystallites nucleated at the surface of the amorphous slab and at the same time of maximizing the vdW attraction energy with the capping layers. In fact, the surface energy is  only very slightly higher for the (111) than for the (001) plane. The theoretical surface energy of the (001) face computed within our framework is 20 meV/{\AA$^2$}, which is higher than the previous estimate by DFT calculations (10.2 meV/{\AA$^2$})\cite{Mandelli} because of the inclusion of vdW interactions. The calculation of the surface energy of the (111) face of the cubic crystal is problematic because a slab with both surfaces terminated by Te is non-stoichiometric. As a reasonable approximation, we can assume that the surface energy of the (111) face of the cubic crystal exposing Te atoms is close to the surface energy of the (0001) face of the hexagonal crystal which can be easily computed from a stoichiometric slab model yielding 16 meV/{\AA$^2$}, which is very close to previous DFT+vdW calculations (14 meV/{\AA$^2$}).\cite{deringerSurface}
The small difference between the surface energy of the two faces 
(4 meV/{\AA$^2$}) might be compensated by the adhesion energy with the capping layers which depends in turn on the surface atomic density, higher for the (001) than for the (111) face. We can then envisage that a small change in density might favor one orientation over the other which also means that the preferential orientation of the crystallites forming at the interface might depend on the type of capping layer and it might then change by 
using, for instance, metal selenides instead of  metal tellurides as spacers in the SLs.

 In the SL at the lower density (SL-LD'), 
  the fully crystallized slab in the simulation at 750 K consists of ten Te layers, as in the original 
 hexagonal crystalline slab, and nine cationic layers with, as expected, no vdW gap and a Te outermost plane on both sides of the slab. 
 This geometry implies that the fraction of vacancies in the cationic sublattice is equal to 1/9,  to be compared with the value of 1/5 in the bulk cubic phase.
 Therefore, in the slab about half of the stoichiometric vacancies of the cubic phase are filled. In the real material, the cubic phase should have a larger concentration of vacancies due to self-doping which turns the system into a degenerate p-type semiconductor.
 Even in the presence of additional non-stoichiometric vacancies responsible for about 2.73 10$^{20}$ holes/cm$^3$ as measured experimentally,
 \cite{holesexp} we expect  in the slab a fraction of vacancies lower than the stoichiometric value which should imply a switch to a n-type conductivity.\cite{Caravati225jpc}

\begin{figure*}[]
\centering
\includegraphics[width=\textwidth, keepaspectratio]{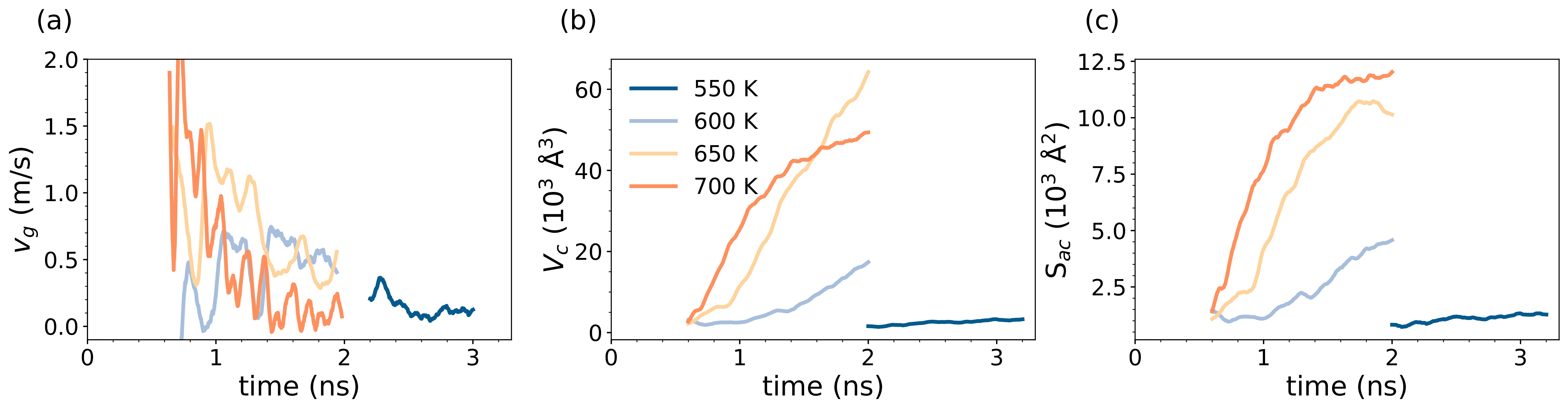}
\caption{(a) Instantaneous crystal growth velocity $v_g$, (b) volume occupied by the crystalline atoms $V_c$ and (c) area of the crystal-amorphous interface $S_{ac}$ as a function of time at the different temperatures in the crystallization of the superlattice configuration (SL-HD). The crystal growth velocity is computed as $v_g=dV_c/dt S_{ac}^{-1}$ as described in Ref. \cite{gst124}.  The  $v_g$ reported in Table \ref{tab:velocities} are obtained by averaging the instantaneous $v_g$ over the time intervals highlighted in Fig. S5 in the ESI\dag.}
\label{fig:v_g_t}
\end{figure*}

\begin{table*}[]
\centering
\small
  \caption{Crystal growth velocities $v_g$ (m/s)  for the superlattice with GST slabs confined by the TiTe$_2$-like capping layers at two different densities, namely the equilibrium density of hexagonal GST with vdW (SL-HD) and the lower equilibrium density without vdW (SL-LD'). The data at 700 K are average with error bars over three independent models.
  The data for the orthorhombic bulk
model at the density of the crystalline hexagonal phase with vdW (bulk-HD)  and at the experimental density of a-GST (bulk-LD) are also reported. The data refer to the calculations with the Voronoi polyhedra (see text). Data in parenthesis for the crystallization in the bulk have been obtained instead from the analysis the radius of the growing nuclei (see text). In case of bulk-HD two values are reported, the first refers to the orthorhombic cell and the second to the larger cubic cell (see text).}
  \label{tab:velocities}
  \begin{tabular*}{0.7\textwidth}{@{\extracolsep{\fill}}ccccc}
    \hline
    &\multicolumn{4}{c}{$v_g$ (m/s)}\\
    \\
Temperature (K) &SL-HD & SL-LD' &bulk-HD &bulk-LD \\
\hline
550 & 0.15 & -    &0.29  (0.35-0.67) &0.70 (0.61) \\
600 & 0.53 & 0.75 &0.70 (1.00-0.98)  &1.63 (1.76) \\
650 & 1.03 & 1.30 &1.40 (2.00-2.40) &(2.45)\\
700 & 1.45$\pm$0.15 & 2.00$\pm$0.3 &2.50 (2.75-2.90) &(2.98)\\
750 & 2.10 & 2.6& 2.20 (2.50-2.40)& (2.72)\\
    \hline
  \end{tabular*}
\end{table*}

Coming now to the crystal growth velocity, due to the presence of a heterogeneous and complex nucleation process, we can not compute $v_{\rm g}$ by using the same scheme reported in Sec. 3.2 for crystallization in the bulk. 
Therefore, for the SL we used the scheme adopted in our previous work on nanoconfined GeTe, \cite{acharya2024} namely $v_{\rm g}$   has been computed 
from the time derivative of the crystalline volume $V_c$  according to the scheme proposed in Ref. \cite{ronneberger2018crystal} as $v_g(t)$=$S_{ac}^{-1}$d$V_c$/dt where  $S_{ac}$ is the area of the crystal-amorphous interface. The crystalline volume $V_c$ is obtained  by summing up the volumes of the Voronoi polyhedra of each crystalline-like atom (excluding the volume of isolated atoms or clusters of less than 28 crystalline-like atoms). $S_{ac}$ is computed as the total area of the faces that are shared by Voronoi polyhedra of amorphous-like and crystalline-like atoms. We used the Voro++ code.\cite{Voronoi+} The data of volumes $V_c$  and areas  $S_{ac}$  were smoothed using a Savitzky–Golay filter with a time window of 10-50 ps for the calculation of growth velocity, similarly to  Ref. \cite{gst124}. 
The instantaneous $v_g$, $V_c$ and $S_{ac}$ as a function of time at the different temperatures is shown in Fig.  \ref{fig:v_g_t}.
The average crystal growth velocities in the SL at high density (SL-HD) are obtained by averaging the instantaneous crystal growth velocity in a time interval of a few hundreds of ps as shown in Fig. S5 in the ESI\dag. 
The resulting $v_g$ are compared in Table \ref{tab:velocities} with those of reference calculations for the bulk at the density of the crystalline hexagonal phase (bulk-HD)  and at the experimental density of a-GST (bulk-LD).

The crystallization in bulk-HD was studied with two models at the same density, namely the 3528-atom orthorhombic cell introduced in Sec. 3.2 and a second 4536-atom cubic cell.
The amorphous models were generated by quenching from the melt in 150 ps. The models were then heated and equilibrated at the target temperature.
Crystal nucleation in the bulk-HD models was observed on a short time scale at 550 in our model and at 600 K in the other. The crystal growth velocity at the higher (lower) temperature was computed by heating (cooling) at the target temperature a configuration from the simulation  with a crystalline nucleus grown up to about 160-170 atoms. This choice is made to ensure that the nucleus remains
overcritical at the higher temperatures.
The crystal growth velocities for the SL simulations at lower density (SL-LD') are also shown in Table \ref{tab:velocities}.
The equivalent of Fig. \ref{fig:v_g_t} and of Fig. S5
for SL-LD', bulk-HD and bulk-LD models are given in Figs. S6-S11 in the ESI\dag.

 The crystal growth velocity for  SL-HD is lower than in the bulk at the same density (see Table \ref{tab:velocities}). The same effect was found for GeTe/TiTe$_2$-like SLs in our previous work\cite{acharya2024} and it was ascribed to the interaction between the growing overcritical nuclei and  the undercritical nuclei that, although they form and disappear, they can hinder the growth of the overcritical ones. Since in the SL geometry, nucleation takes place at the surface, the different nucleation centers are on average closer in the slab than in the bulk. The same explanation can hold here for the SL-HD model, especially at the lowest temperatures. The SL-LD' and bulk-LD do not have the same density and therefore a direct comparison is not possible, but it is confirmed that $v_{\rm g}$ increases by decreasing density because of a higher diffusion coefficient. 

We remark that the atomic mobility in the xy plane (perpendicular to the z axis of the SL) is very similar in the SL and in the bulk  at the same density, as shown by the diffusion coefficient $D$ in Fig. S12 and 
Table S2  in the ESI\dag, and therefore the difference in  $v_{\rm g}$ cannot be ascribed to a change in $D$ in the kinetic prefactor $u_{kin}$ of the WF formula (Eq. \ref{WF})).  $D$ is computed from   the Einstein relation and  the MSD reported in Fig. S13 in the ESI\dag.

\begin{figure}[h!]
\centering
\includegraphics[width=0.45\textwidth, keepaspectratio]{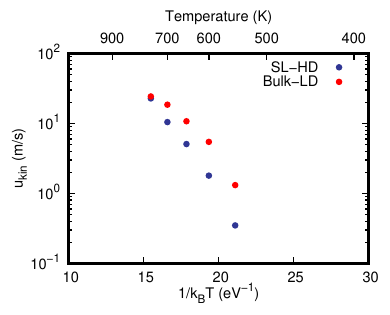}
\caption{$u_{kin}$  (Eq. \ref{WF}) as a function of temperature for the superlattice SL-HD, and  for the bulk at the experimental density of the amorphous phase (bulk-LD, see text). $u_{kin}$ is extracted from $v_{\rm g}$ and the application of the WF formula (Eq. \ref{WF}). The term $\Delta\mu_S$ is included for the bulk only.}
\label{FSC}
\end{figure}

Regarding crystallization of GST in confined geometry,
 a fragile-to-strong crossover (FSC) was inferred in Ref. \cite{chenW} from differential scanning calorimetry (DSC) on a 7 nm film of GST confined by W capping layers. The DSC traces were fitted by the  Johnson–Mehl–Avrami  model for the growth process  of already formed nuclei,\cite{chenW} which implies that the FSC should arise from the kinetic prefactor $u_{kin}$ of the WF formula (Eq. \ref{WF}) for $v_{\rm g}$.  $u_{kin}$ follows a super-Arrhenius behavior
above the strong-fragile crossover temperature T$_{\rm fs}$ and a simple
Arrhenius behavior below T$_{\rm fs}$, which
 was estimated as 1.25 T$_g$=473 K for the 7 nm film capped by W,\cite{chenW}  which is 
 below the temperature at which we see nucleation and growth on our simulation time scale. 
 Indeed, we do not see any clear sign of a crossover in $u_{kin}$ down to 500 K, neither for the bulk at the experimental density (bulk-LD) nor for the SL-HD system (3 nm thick GST) as shown in Fig. \ref{FSC}. $u_{kin}$ is 
extracted from  our $v_{\rm g}$ and the application of WF formula (Eq. \ref{WF}). We remark that a recent measurement of the crystal growth velocity
below about 475 K by nanocalorimetry in GST films 10-40 nm thick,\cite{Zhao2022} revealed  a strong behavior as well; extrapolation of this data at high temperatures leads the authors to estimate T$_{\rm fs}$ at about 680 K which is not consistent with our findings (see also Fig.  \ref{diffusion}). The lower value of $u_{kin}$ of the SL with respect to the bulk is due in part to the different density and in part to the fact that $u_{kin}$  for the SL should also embody the 
effect of the interaction with undercritical nuclei at the surface of the slab, mentioned above, via an effective diffusion coefficient $D_{eff}$ in Eq. \ref{WF}.

Overall, we conclude that the confinement has not a dramatic effect on the crystallization kinetic of our models of GST/TiTe$_2$-like SLs. In fact, $v_g$ of the confined GST slab at the temperature of maximal crystallization speed is comparable (within a factor two at most) to that of the bulk.
This feature makes GST/TiTe$_2$  SLs a viable candidate for applications in neuromorphic computing with improved data retention.

\section{Conclusions}
In summary, we have exploited a recently devised NN potential for GST  to study the effect on nanoconfinement on the 
crystallization kinetics of its amorphous phase. We considered
ultrathin (3.14 nm) slabs of GST confined by capping layers aimed at mimicking the TiTe$_2$ spacers in GST/TiTe$_2$ superlattices,  analog to the Sb$_2$Te$_3$/TiTe$_2$ heterostructure proposed in Ref. \cite{DingPCH} for applications in neuromorphic devices. The replacement of Sb$_2$Te$_3$ by GST would improve the data retention of the device due to higher crystallization temperature of GST with respect to   Sb$_2$Te$_3$. 
The simulations show that nanoconfinement leads to a decrease of the crystal growth velocity  with respect to the bulk amorphous phase which is, however,  rather minor for the perspective application in neuromorphic devices.  On the other hand, we also observed an increase in the nucleation rate with respect to the bulk due to heterogeneous nucleation.
In conclusions, MD simulations support the idea of investigating GST/TiTe$_2$ superlattices for applications in neuromorphic devices with improved data retention as also proposed in Ref. \cite{mazzarello2024}.

\section*{Conflicts of interest}
There are no conflicts of interest to declare.

\section*{Code availability}
LAMMPS, and DeePMD are free and open source codes available at https://lammps.sandia.gov and http://www.deepmd.org, respectively. 

\section*{Data Availability Statement}
The trajectories of the crystallization process of the different models will be available as additional materials in the Materials Cloud repository. The neural network potential generated in O. Abou El Kheir et al. npj Comput. Mater. 2024, 10, {\bf 33}, is available in the Materials Cloud  repository at https://doi.org/10.24435/materialscloud:a8-45.

\section*{Author Contributions}
M.B conceptualized the work and wrote the original draft of the paper.
The simulations were performed by D.A. and S.M.. All authors contributed to the analysis of the data and approved the final version of the paper.
\section*{Acknowledgments}
We acknowledge financial support from the PRIN 2020 project “Neuromorphic devices based on chalcogenide heterostructures” No. 20203K2T7F funded by the Italian Ministry for University and Research (MUR).
The project has received funding also from European Union Next-Generation-EU  through the Italian Ministry of University and Research under PNRR M4C2I1.4 ICSC 371 Centro Nazionale di Ricerca in High Performance Computing, Big Data and Quantum 372 Computing (Grant No. CN00000013). 
We acknowledge the CINECA award under the ISCRA initiative, for the availability of high-performance computing resources and support.
We gratefully thank R. Mazzarello
for discussions and information.

\bibliographystyle{rsc}
\bibliography{refNew}

\onecolumngrid
\newpage
\noindent{\large\textbf{Simulation of the crystallization kinetics of  Ge$_2$Sb$_2$Te$_5$ nanoconfined in superlattice geometries for phase change memories - Supplementary Information}}
\setcounter{figure}{0}    
\setcounter{table}{0}   

\begin{table}[h!]
  \renewcommand\tablename{Table~S$\!\!$}\small
  \caption{Average coordination number for different pairs of atoms computed from the partial pair correlation functions for models of amorphous GST generated with and without  (from Ref. \cite{npjOmar}) vdW  corrections, both at the experimental density of the amorphous phase of 0.0309 atom/\AA$^3$ (bulk-LD).}
  \label{tab:sTablecoorBulk}
  \begin{tabular*}{0.48\textwidth}{@{\extracolsep{\fill}}rcccc}
    \hline
    & & & with vdW & without vdW\\
    \hline
&Ge &with Ge &0.33 &0.33\\
& &with Sb &  0.30&0.33\\
& &with Te & 3.35&3.53\\
& &Total & 3.99&4.19\\
&Sb &with Ge & 0.30&0.33\\
& &with Sb & 0.55&0.51\\
& &with Te &3.20 &3.64\\
& &Total & 4.06&4.46\\
&Te &with Ge & 1.34&1.41\\
& &with Sb & 1.28&1.45\\
& &with Te &0.31 &0.29\\
& &Total & 2.93&3.16\\
    \hline
  \end{tabular*}
\end{table}

\clearpage
\begin{figure}[h!]
\renewcommand\figurename{Figure~S$\!\!$}
\centering{
\includegraphics[width=0.3\linewidth,keepaspectratio]{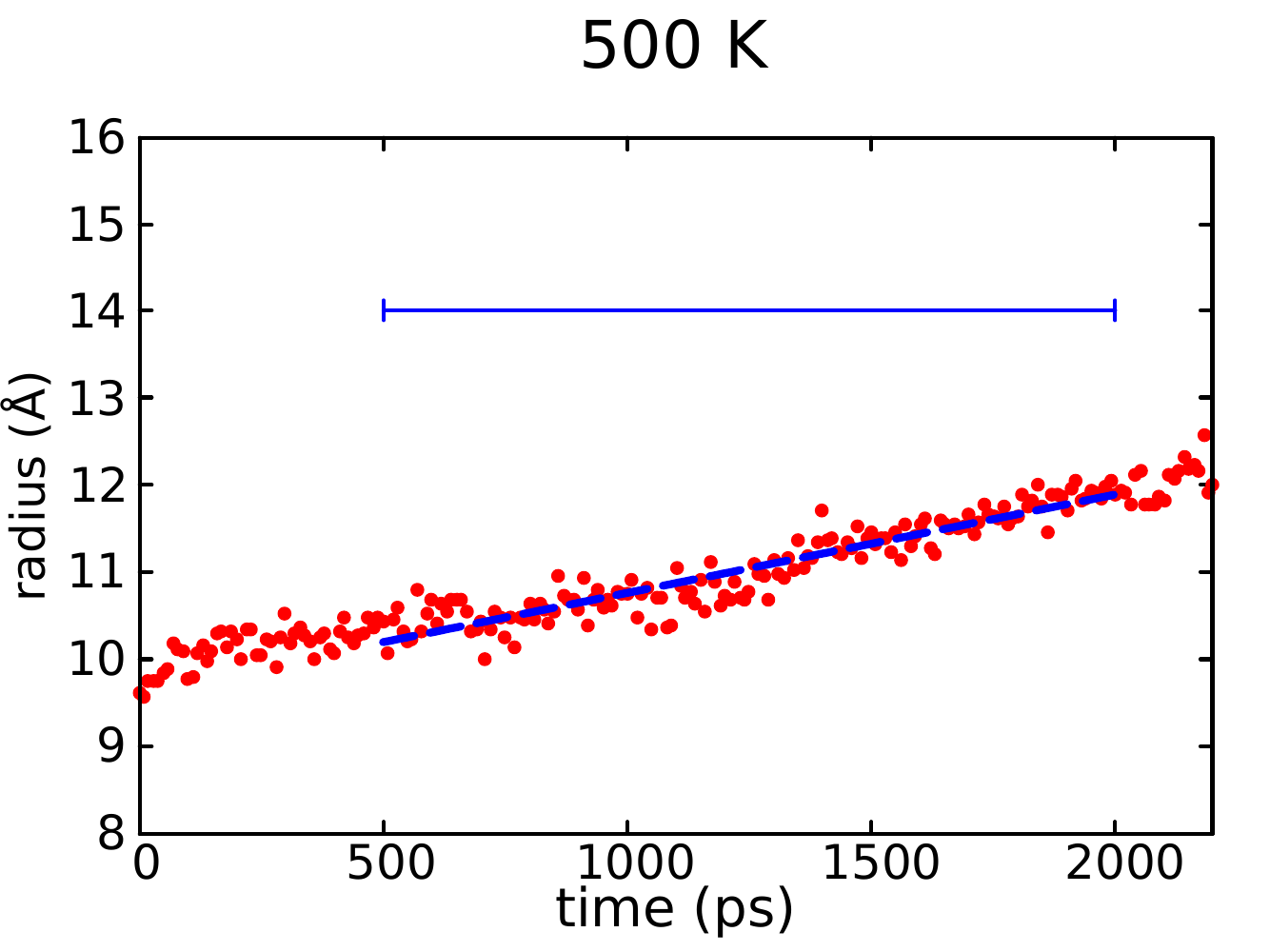}
\includegraphics[width=0.3\linewidth,keepaspectratio]{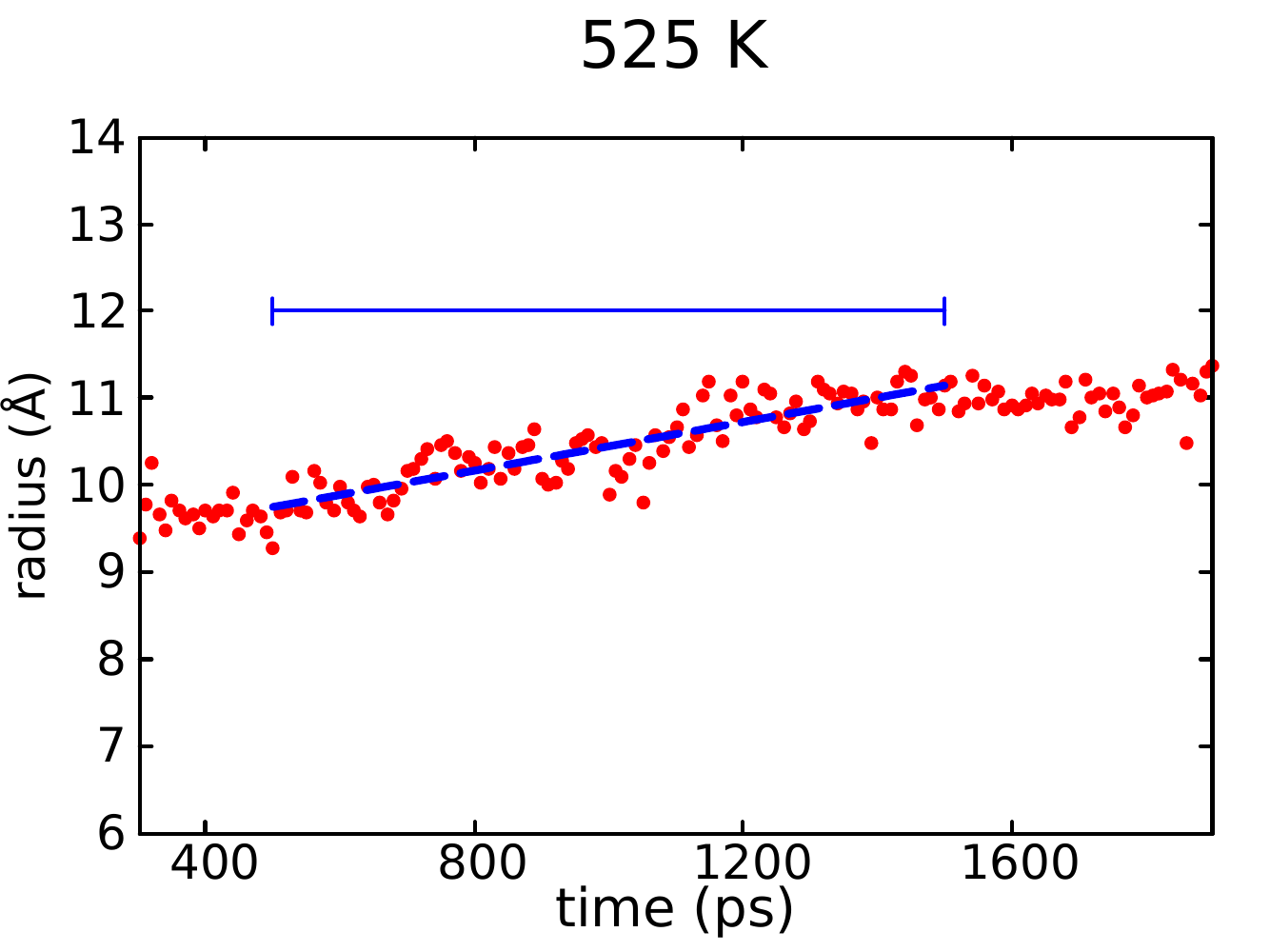}
\includegraphics[width=0.3\linewidth,keepaspectratio]{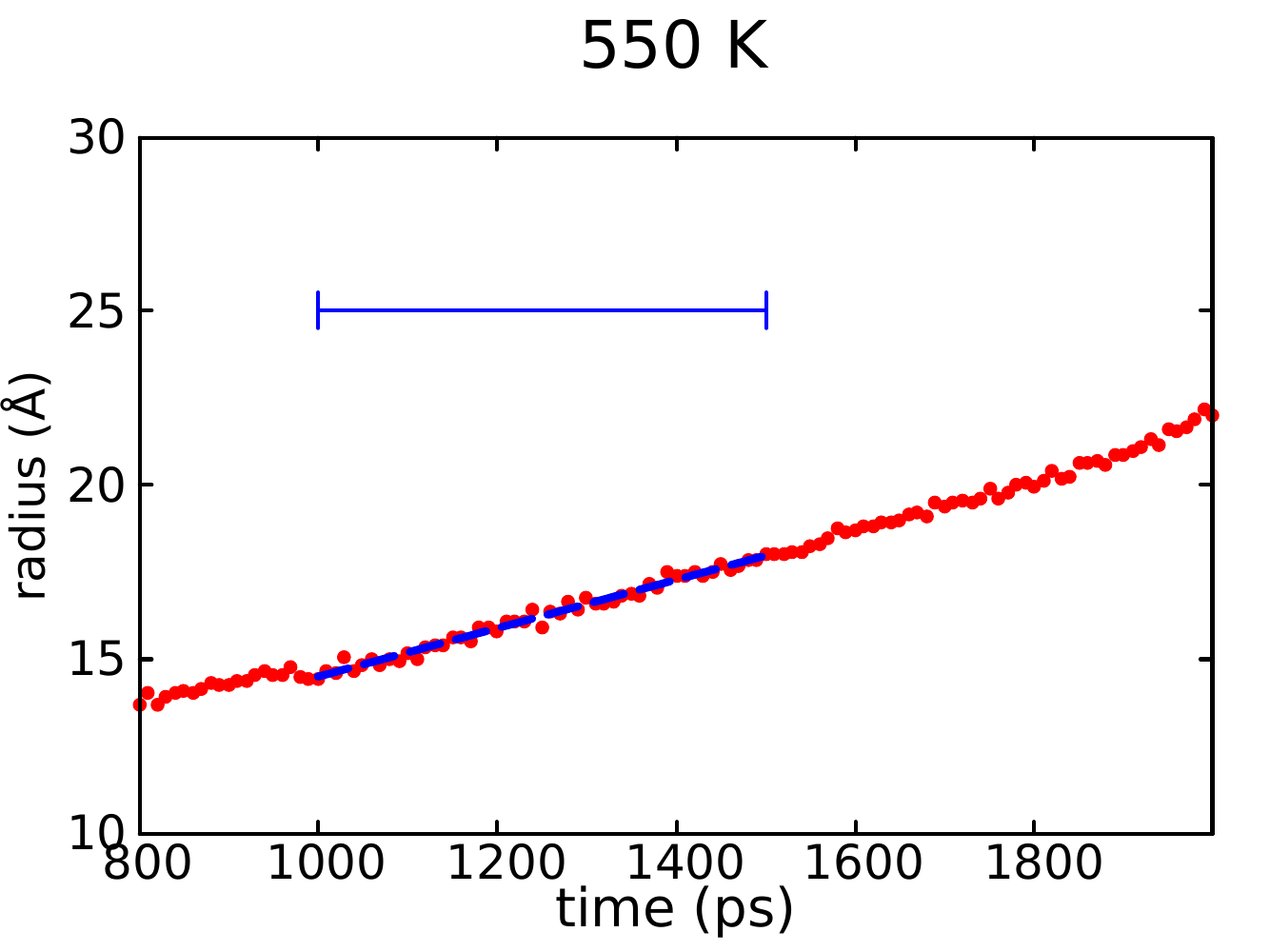}
\includegraphics[width=0.3\linewidth,keepaspectratio]{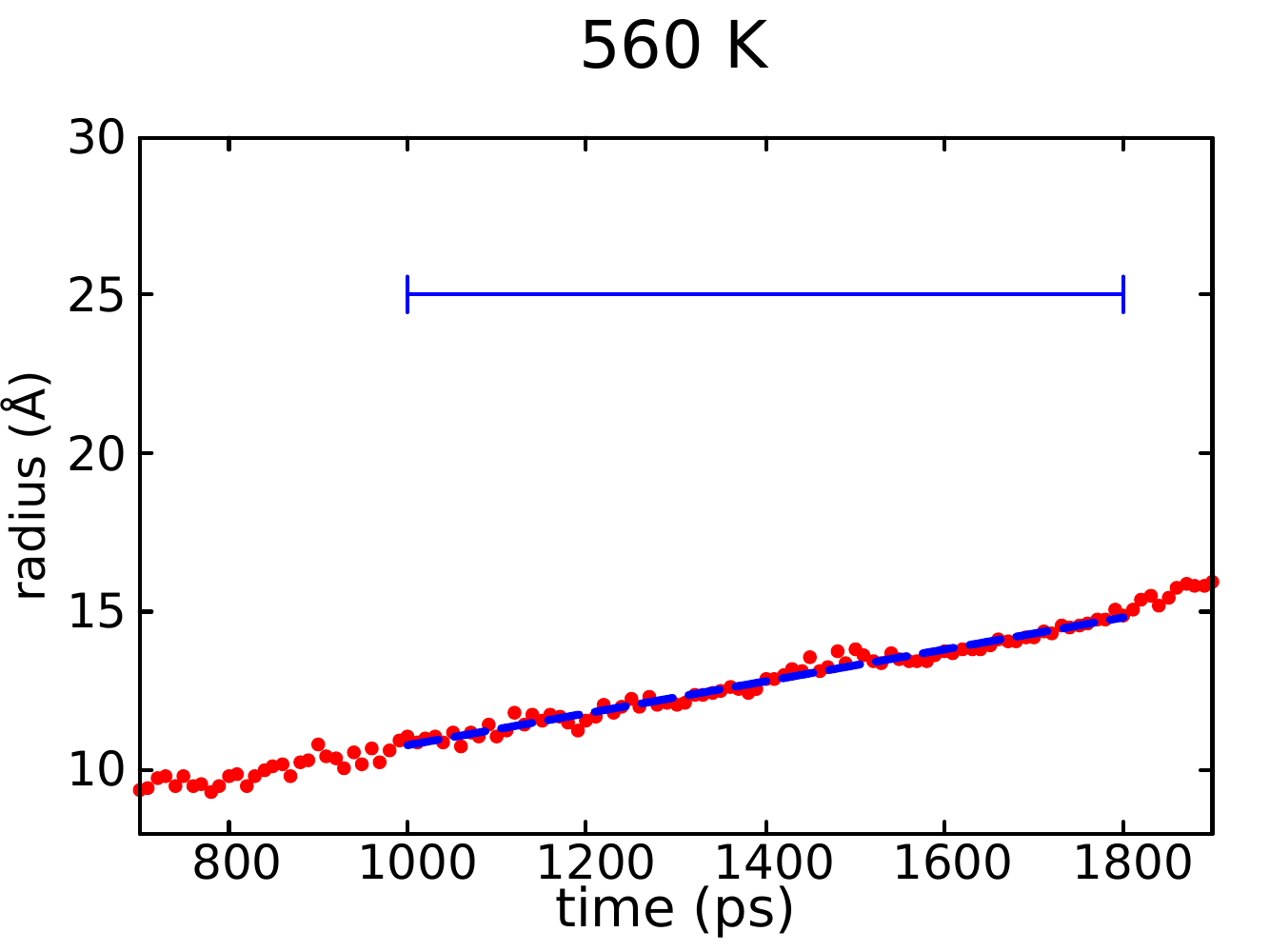}
\includegraphics[width=0.3\linewidth,keepaspectratio]{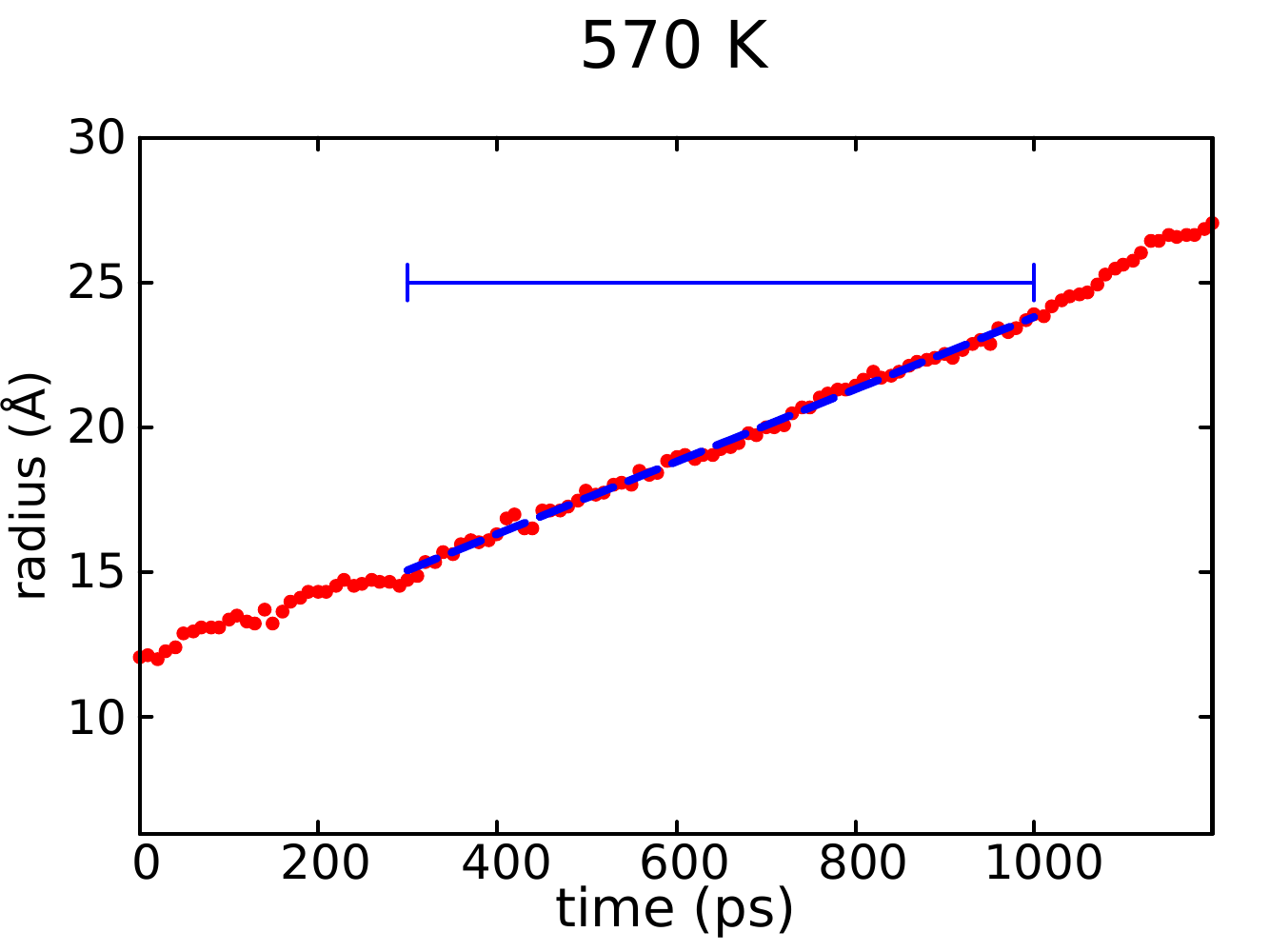}
\includegraphics[width=0.3\linewidth,keepaspectratio]{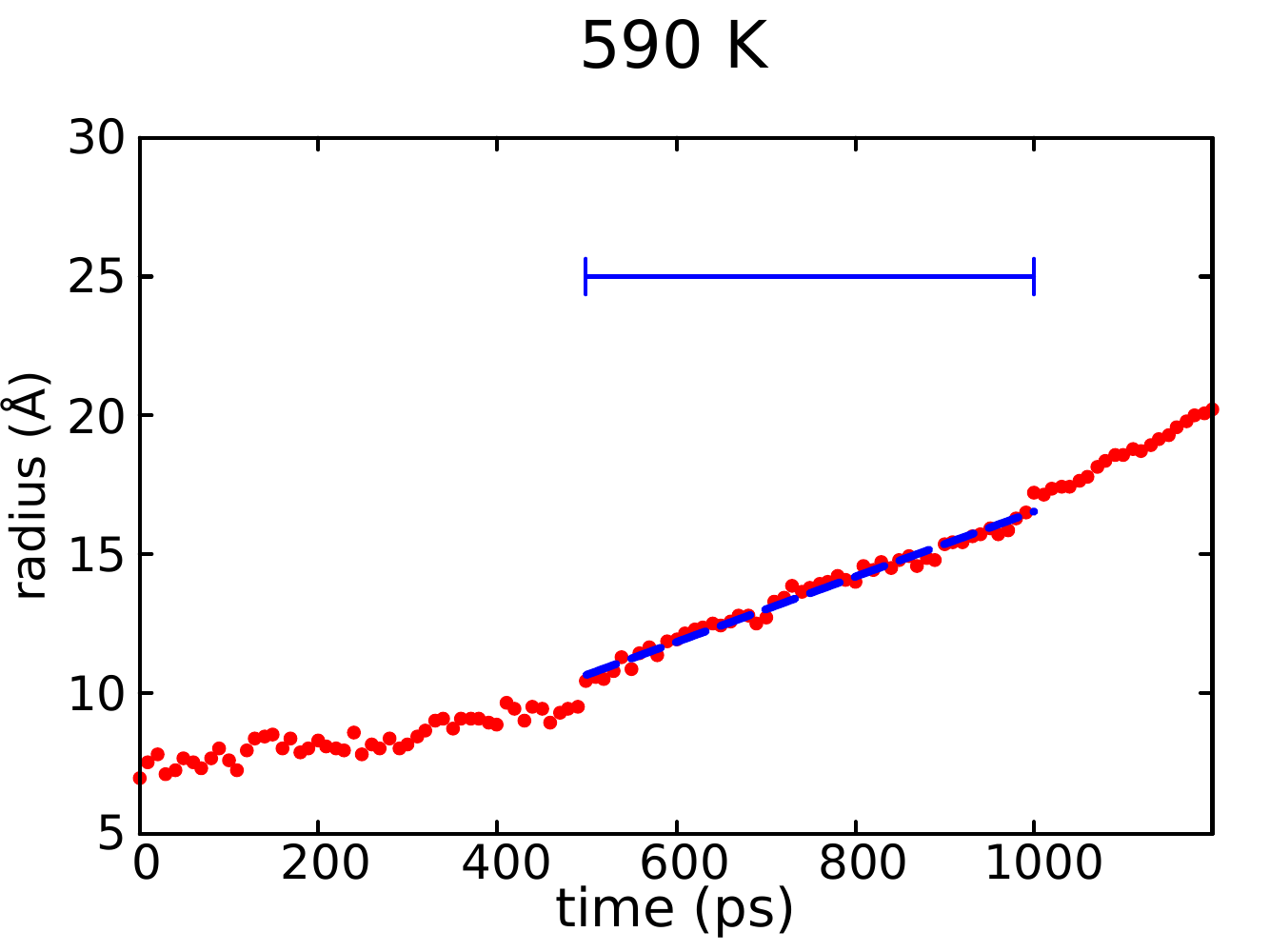}
\includegraphics[width=0.3\linewidth,keepaspectratio]{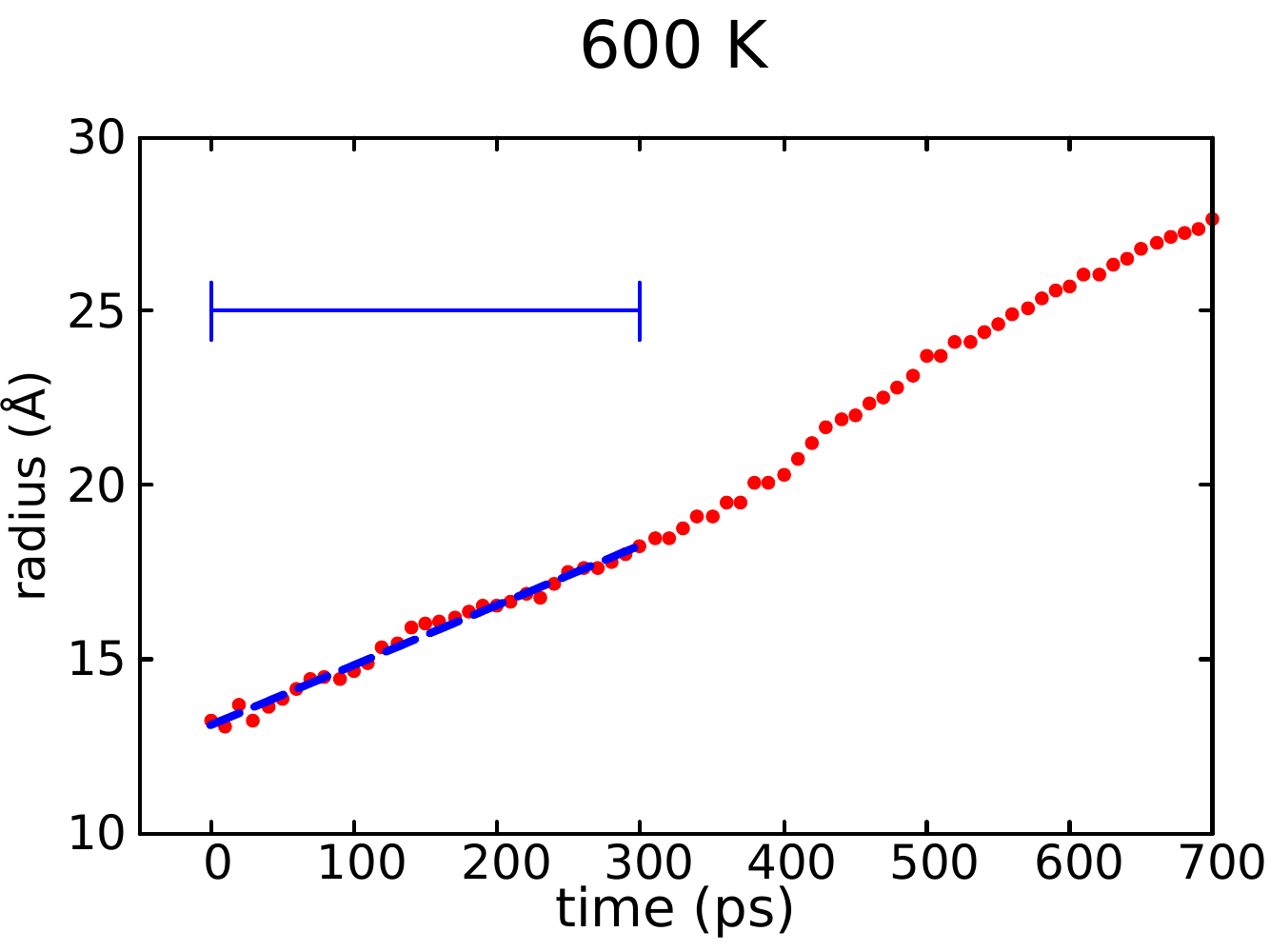}
\includegraphics[width=0.3\linewidth,keepaspectratio]{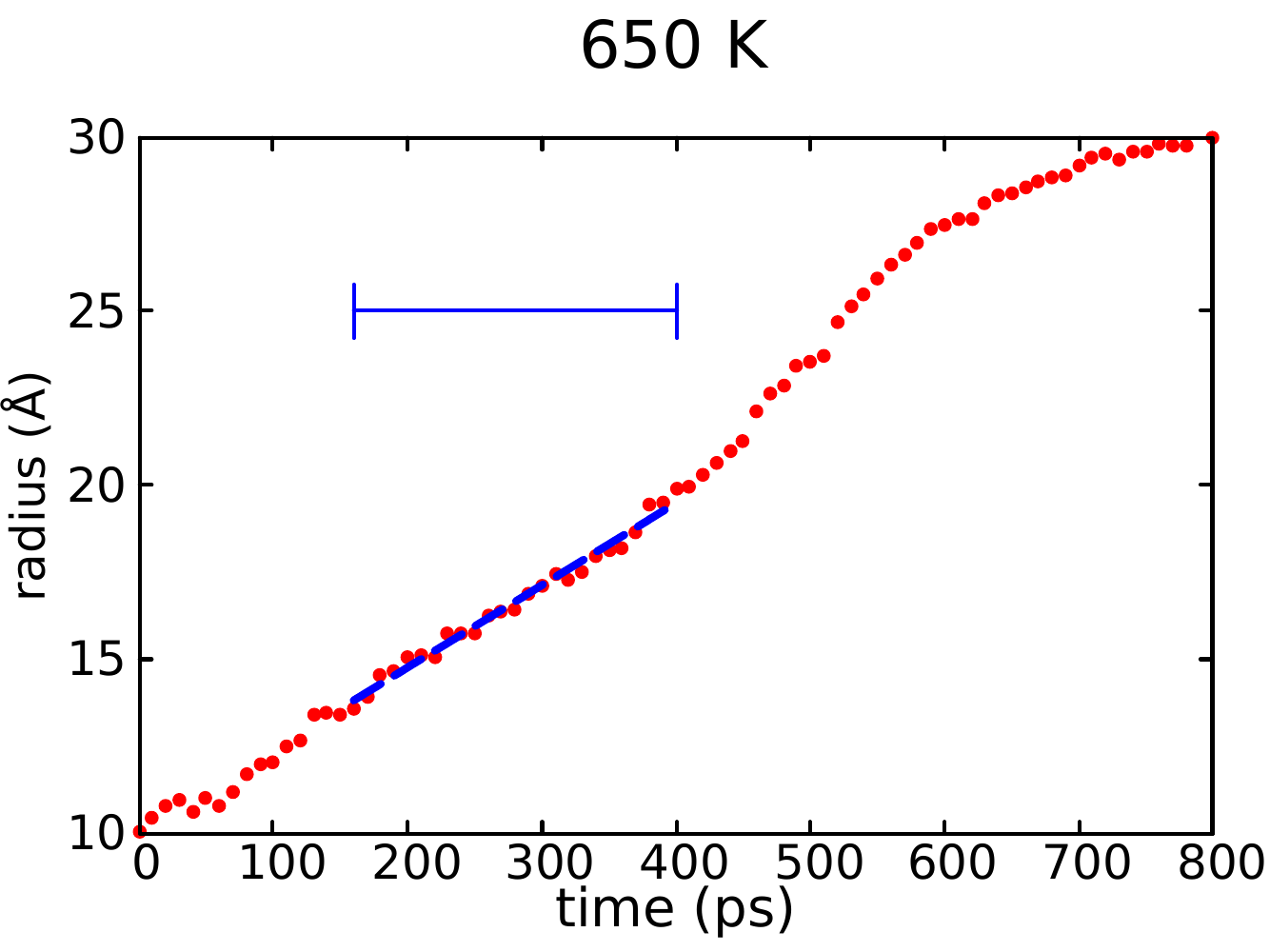}
\includegraphics[width=0.3\linewidth,keepaspectratio]{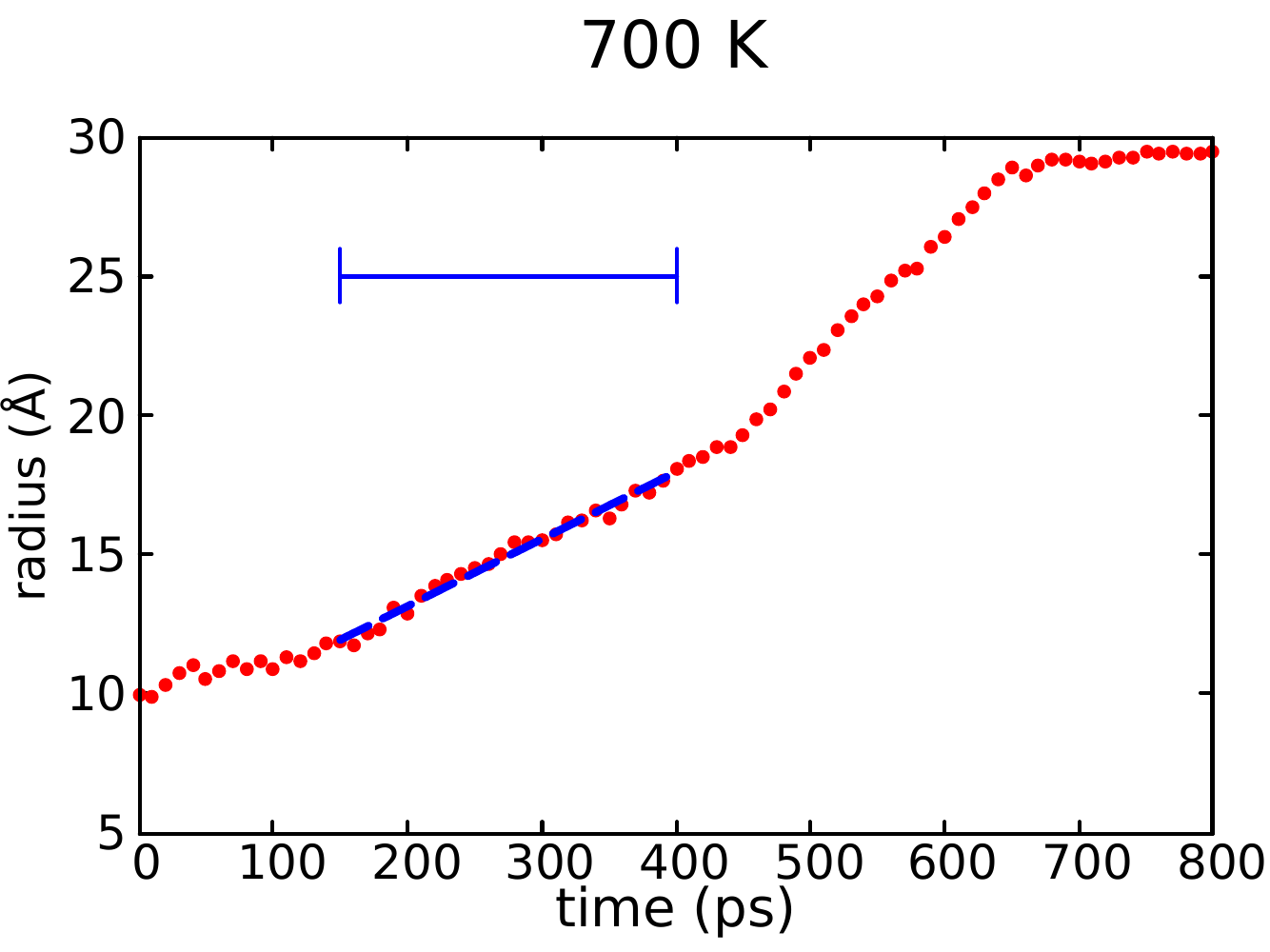}
\includegraphics[width=0.3\linewidth,keepaspectratio]{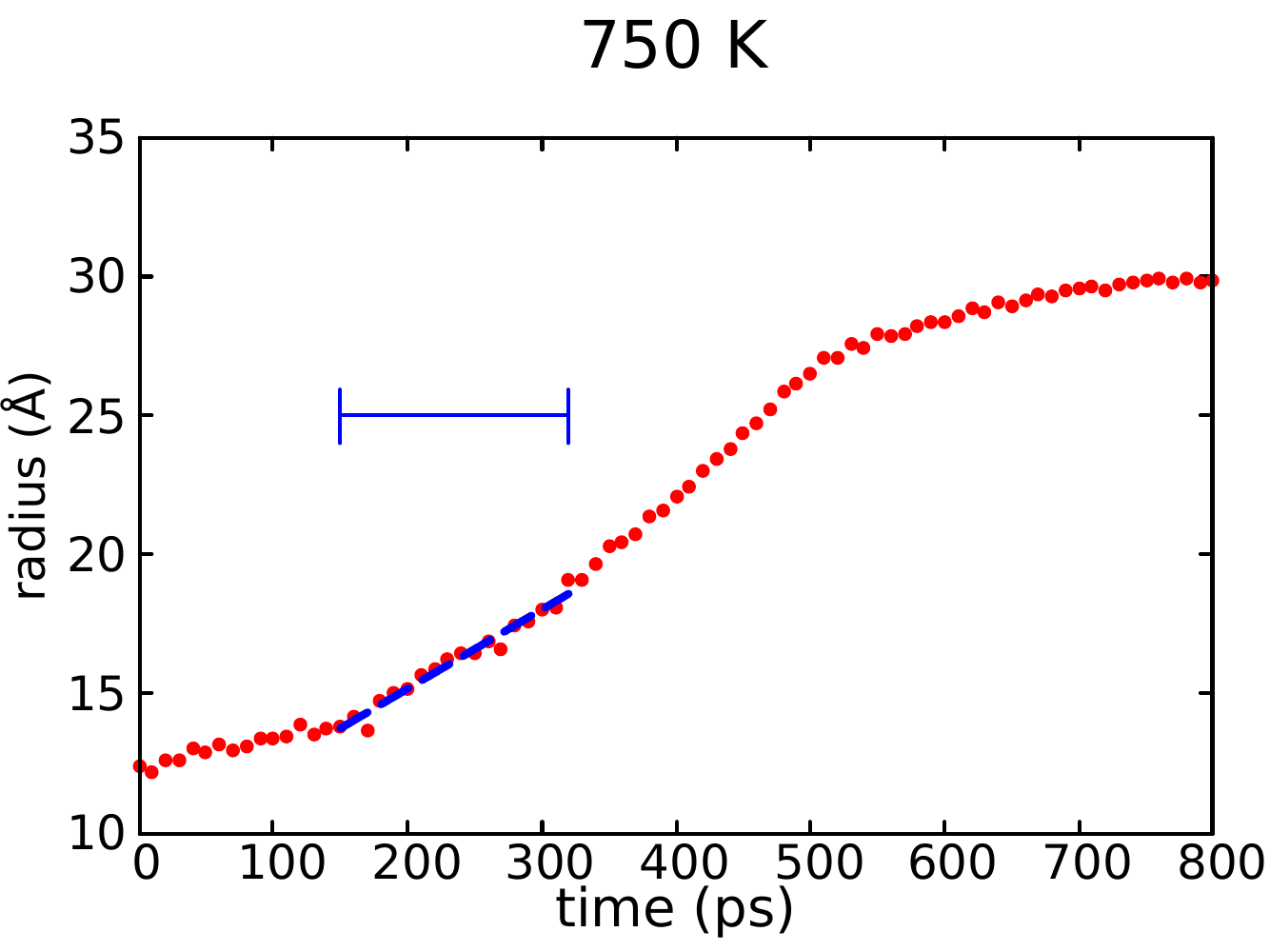}
\includegraphics[width=0.3\linewidth,keepaspectratio]{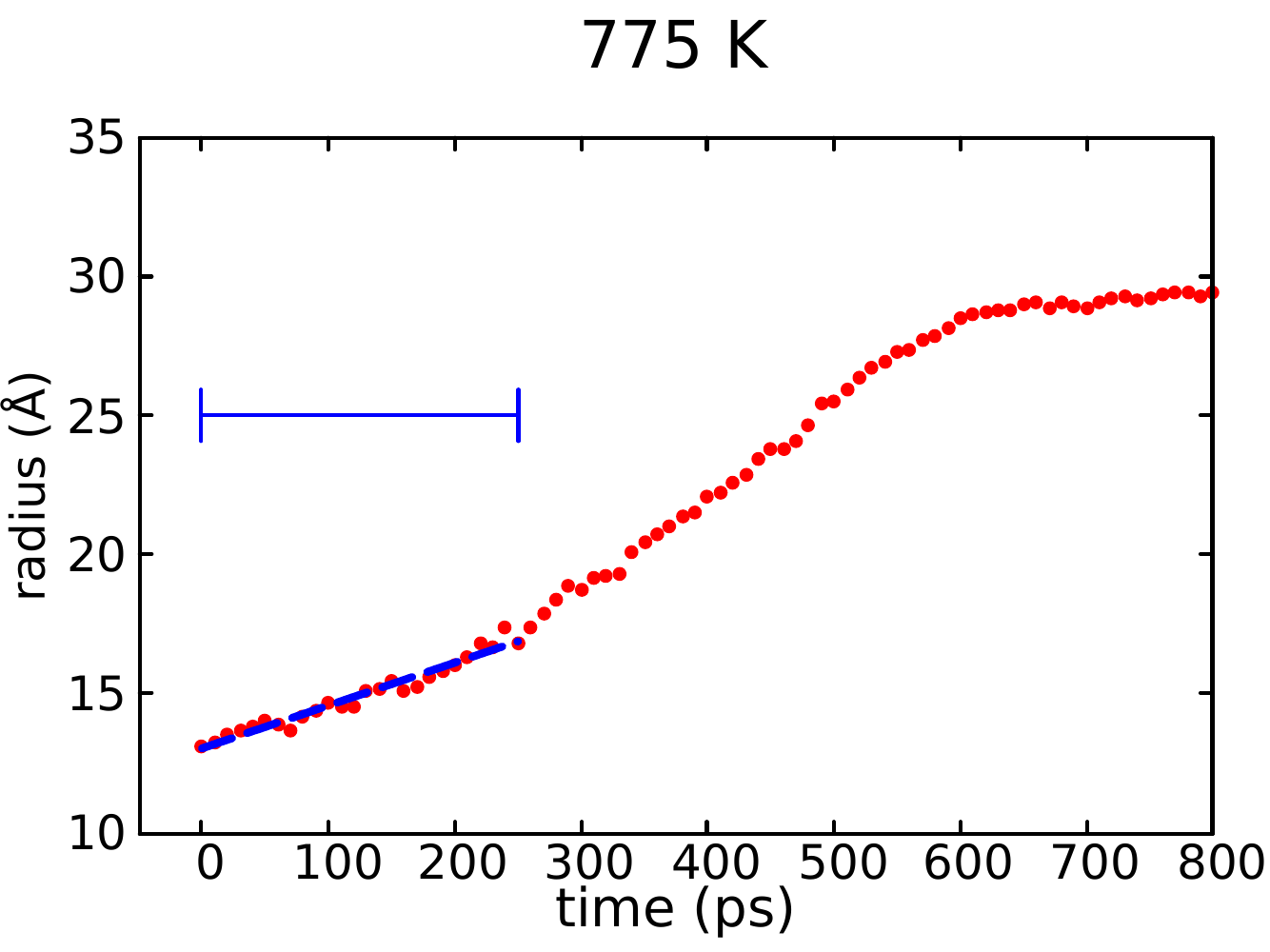}
\includegraphics[width=0.3\linewidth,keepaspectratio]{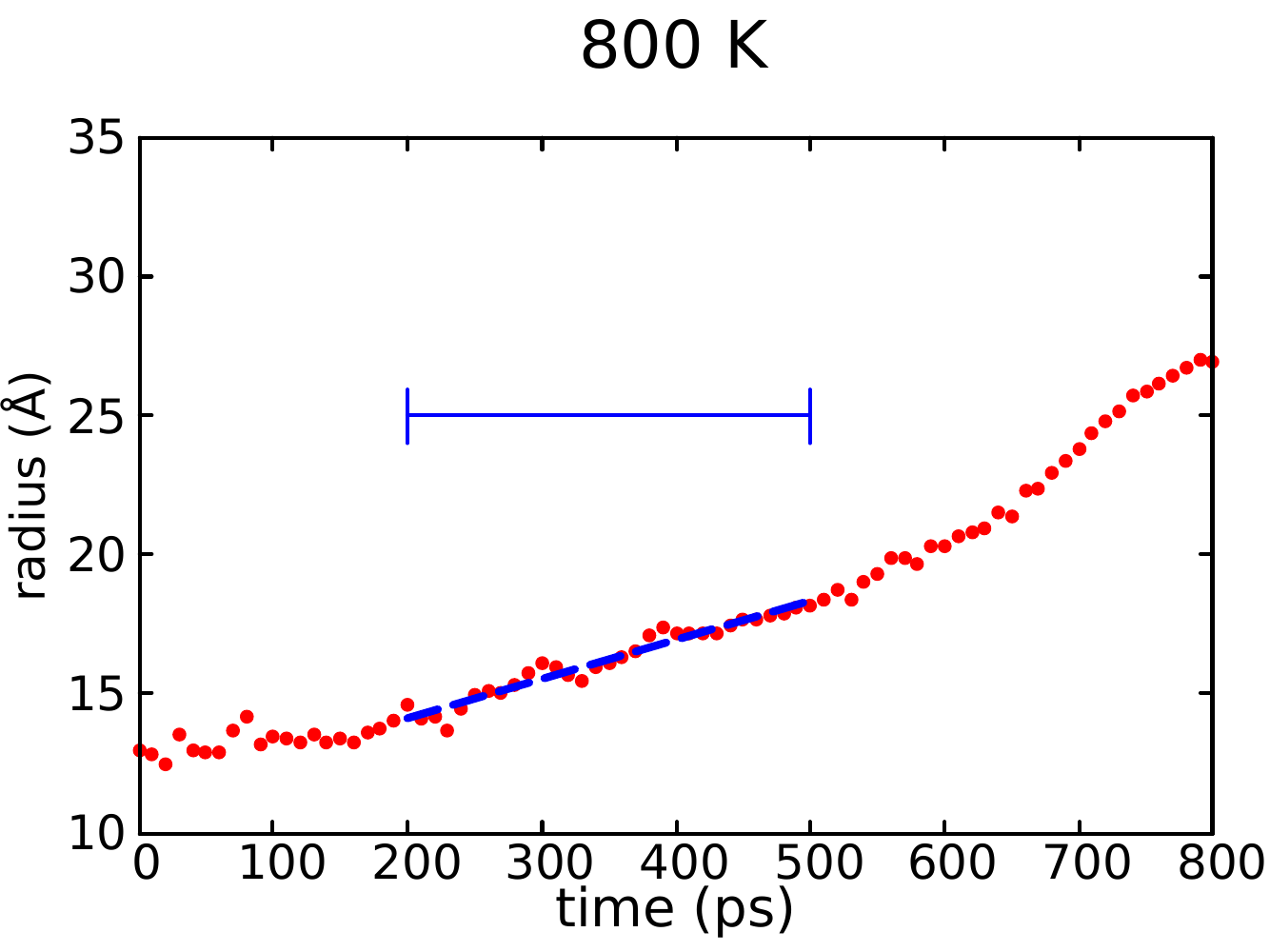}}
\caption{The evolution of the radius of the crystalline nuclei as a function of time in the NN+vdW simulations of the bulk at different temperatures and  at the experimental density (bulk-LD). The linear fit in the range highlighted by horizontal bars yields the crystal growth velocity (see Table 3 in the article).
}
\label{fig:sradius}
\end{figure}

\newpage

\begin{figure}[h]
  \renewcommand\figurename{Figure~S$\!\!$}
\centering
{\includegraphics[height=12cm]{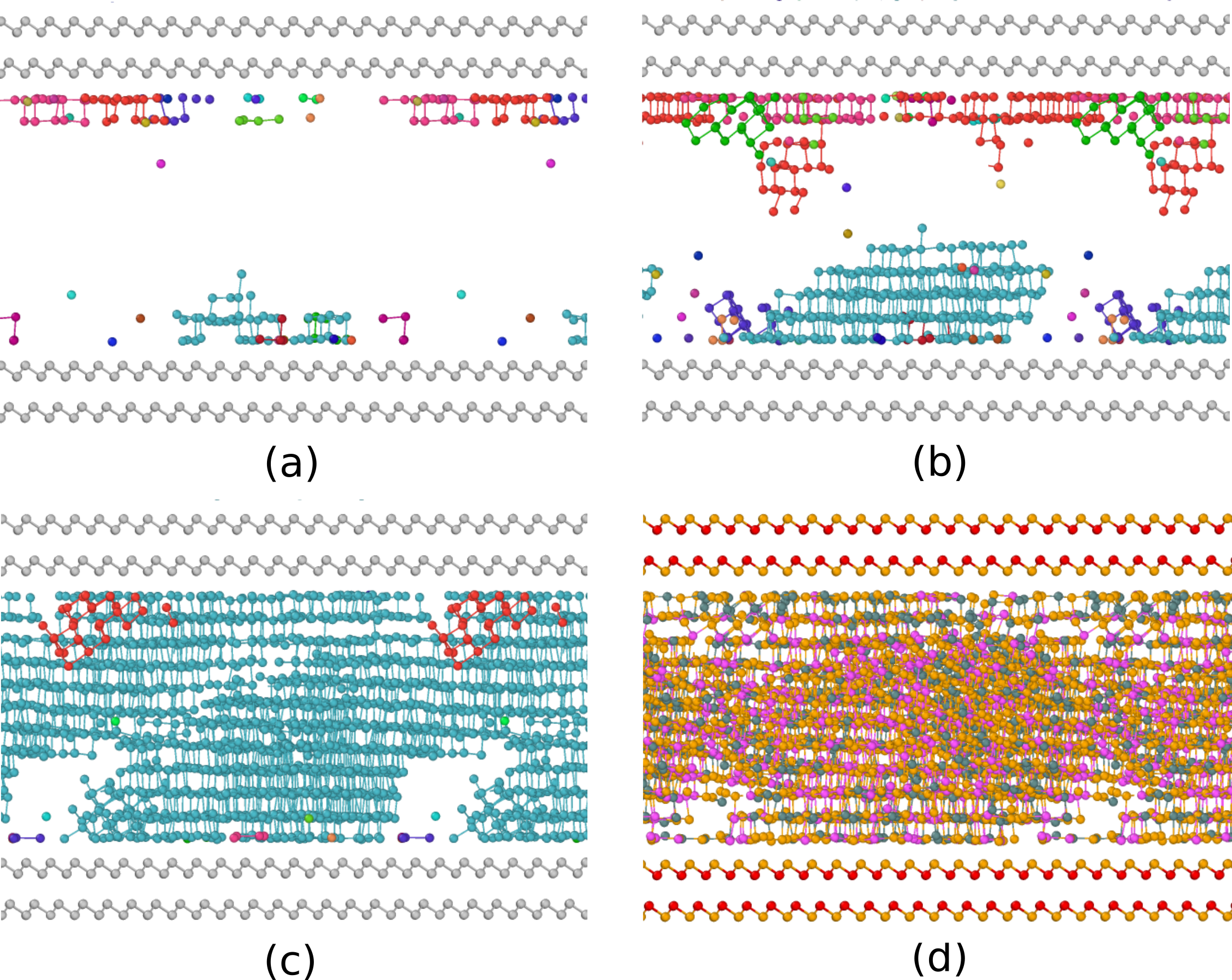}}
\\

\caption{Simulation of the crystallization of a 3528-atom slab of amorphous GST capped by bilayers mimicking confinement by TiTe$_2$ in GST/TiTe$_2$-like  SL (SL-HD). Snapshots at different times at 650 K are shown for (a) 0.5 ns (b) 1 ns, and  (c) 1.5 ns. Only crystalline atoms, identified by the $Q_4^{dot}$ order parameter (see article), are shown. Different crystalline nuclei  have different colors. d) Final configuration after 2 ns. The color code is the same of Fig. 1 of the article. }
\label{fig:sCryst650}
\end{figure}

\begin{figure}[h]
  \renewcommand\figurename{Figure~S$\!\!$}
\centering
{\includegraphics[height=12cm]{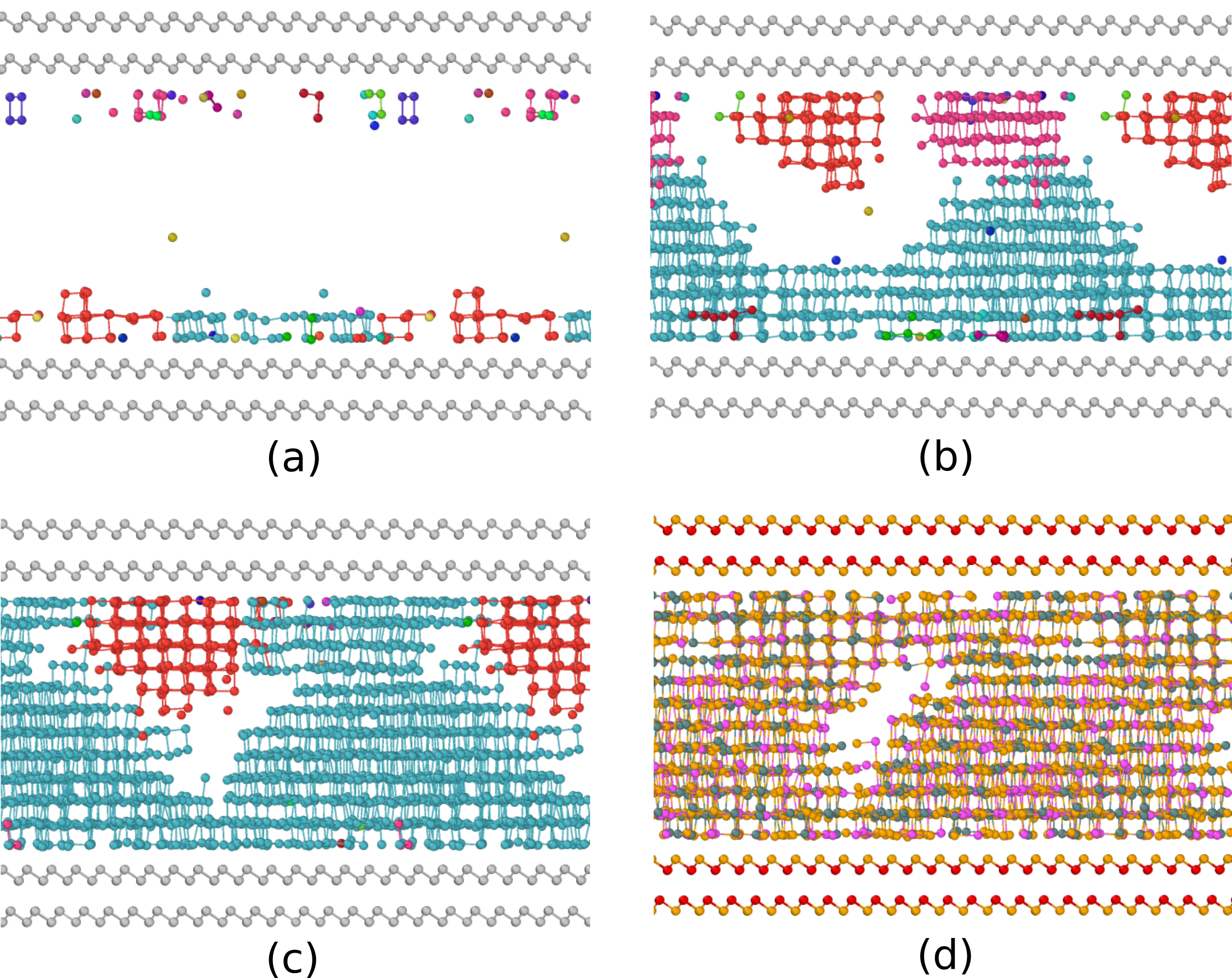}}
\\
\caption{Simulation of the crystallization of a 3528-atom slab of amorphous GST capped by bilayers mimicking confinement by TiTe$_2$ in GST/TiTe$_2$-like  SL (SL-HD). Snapshots at different times at 700 K are shown for (a) 0.5 ns (b) 1 ns, and  (c) 1.5 ns. Only crystalline atoms, identified by the $Q_4^{dot}$ order parameter (see article), are shown. Different crystalline nuclei  have different colors. d) Final configuration after 2 ns. The color code is the same of Fig. 1 of the article.}
\label{fig:sCryst700}
\end{figure}
\clearpage
\begin{figure}[h]
\centering
  \renewcommand\figurename{Figure~S$\!\!$}
 {\includegraphics[height=6cm]{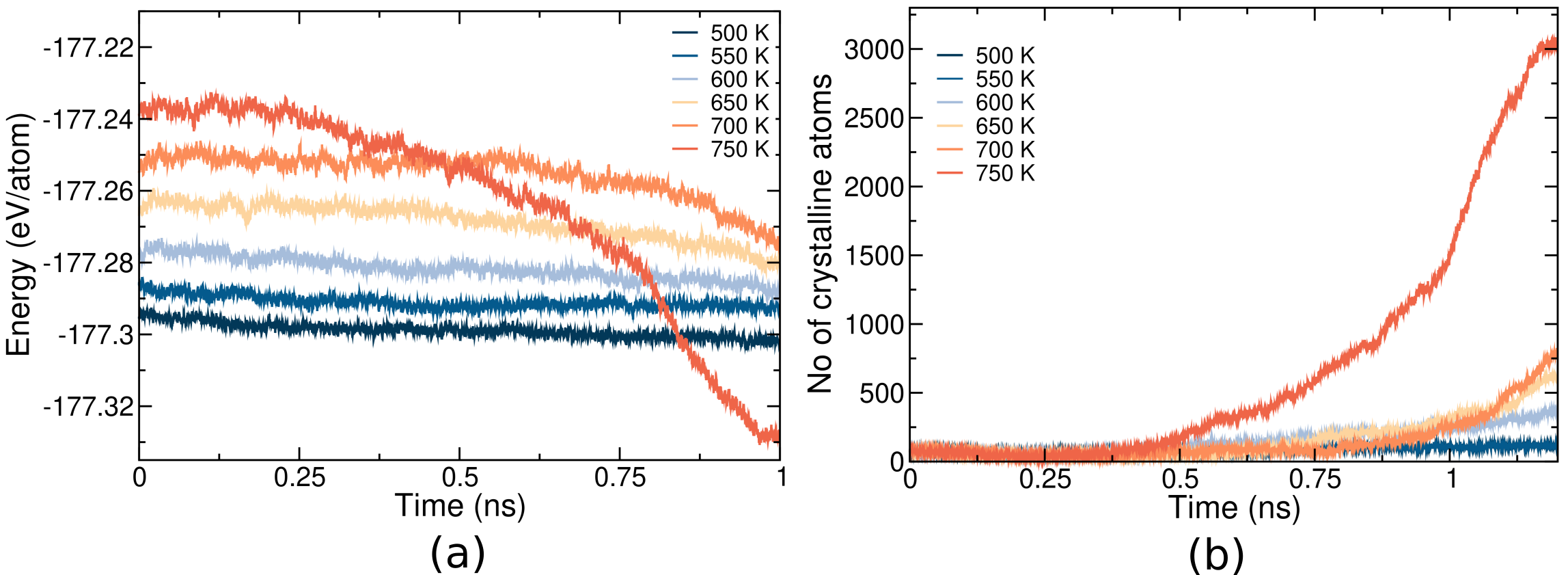}}
\caption{Crystallization of amorphous GST capped by bilayers mimicking confinement by TiTe$_2$ in GST/TiTe$_2$ SL at lower density (SL-LD', see text)
at different temperatures.  a) Potential energy and b) number of crystalline atoms as a function of time.}
\label{fig:sinternalenergyLD'}
\end{figure}

\begin{figure}[h]
  \renewcommand\figurename{Figure~S$\!\!$}
    \centering
\includegraphics[width=\textwidth,keepaspectratio]{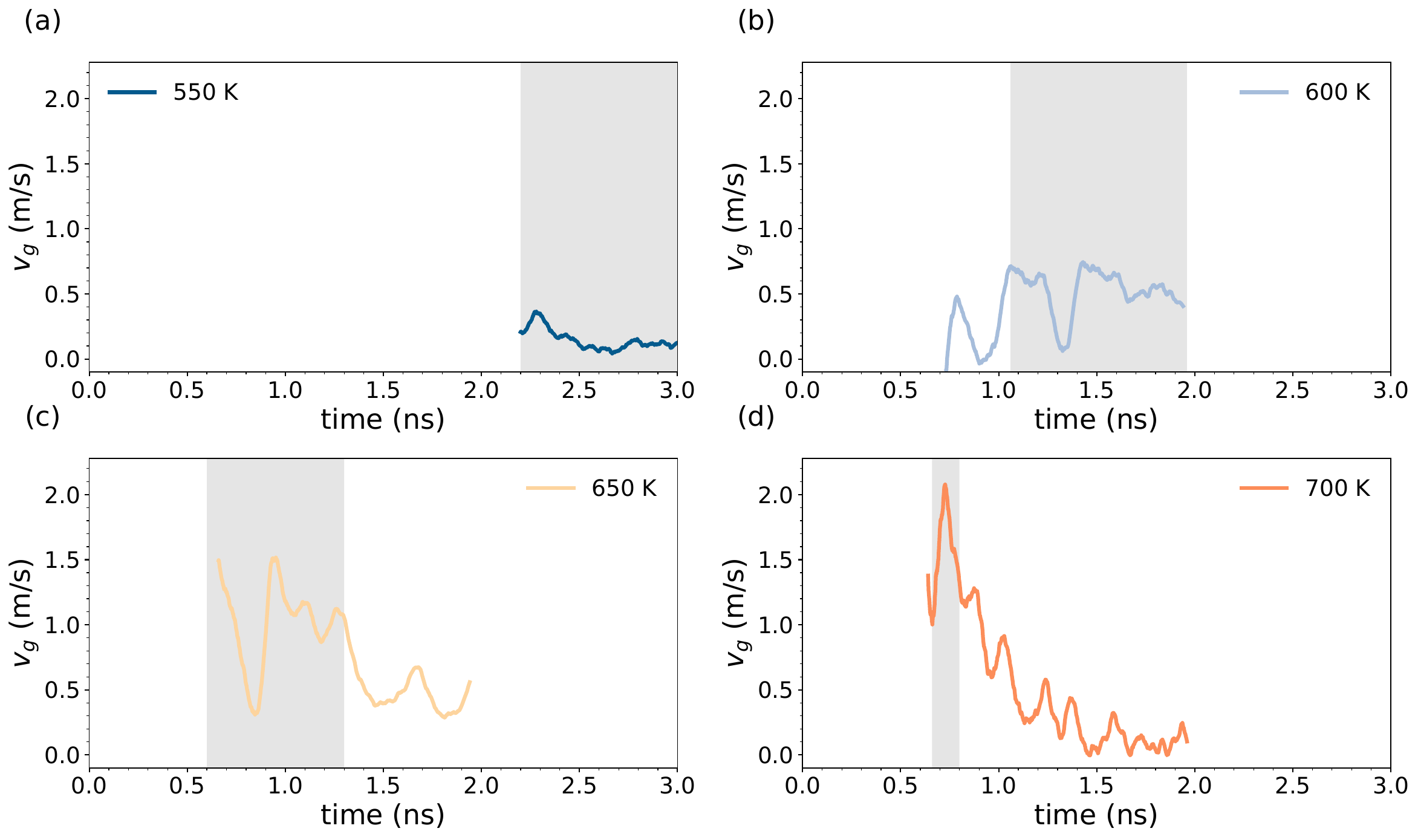}
\includegraphics[width=0.5\textwidth,keepaspectratio]{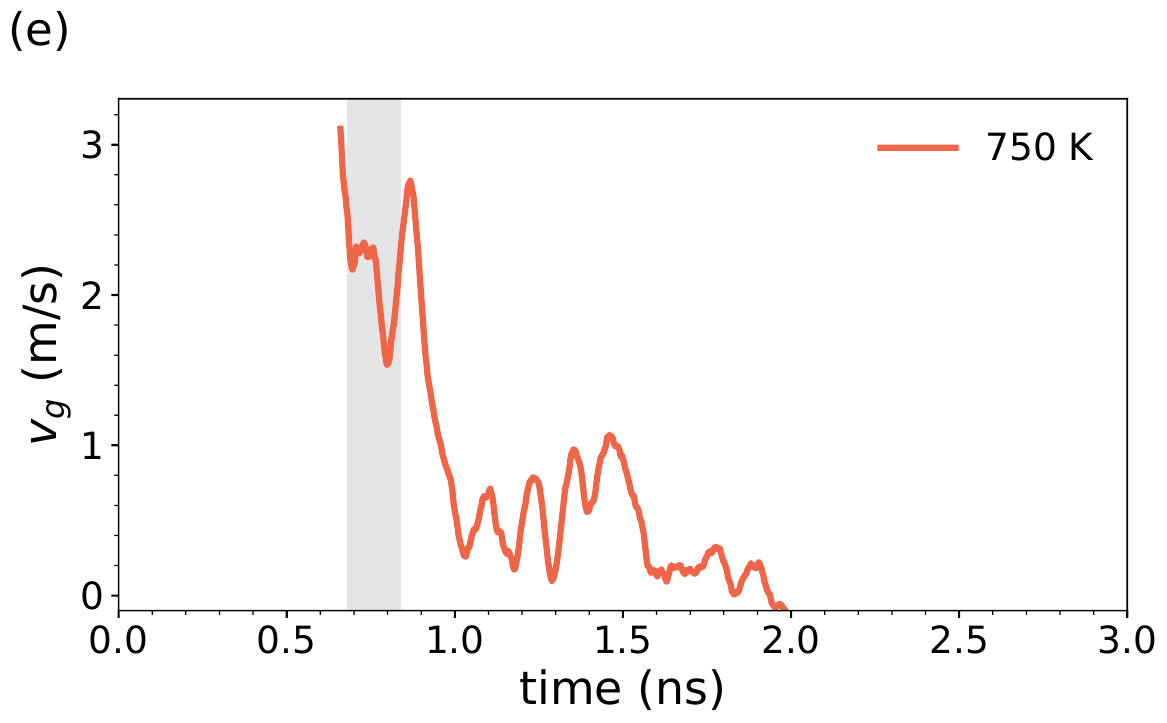}

    \caption{a)-d) Instantaneous crystal growth velocity ($v_g$) as a function of time  at different temperatures for
    the SL-HD model. The region highlighted in gray corresponds to the time interval over which we estimated the average crystal growth velocities reported in Table 3 in the article.}
 \label{fig:smy_label}
\end{figure}

\begin{figure}[h]
  \renewcommand\figurename{Figure~S$\!\!$}
    \centering
\includegraphics[width=\textwidth,keepaspectratio]{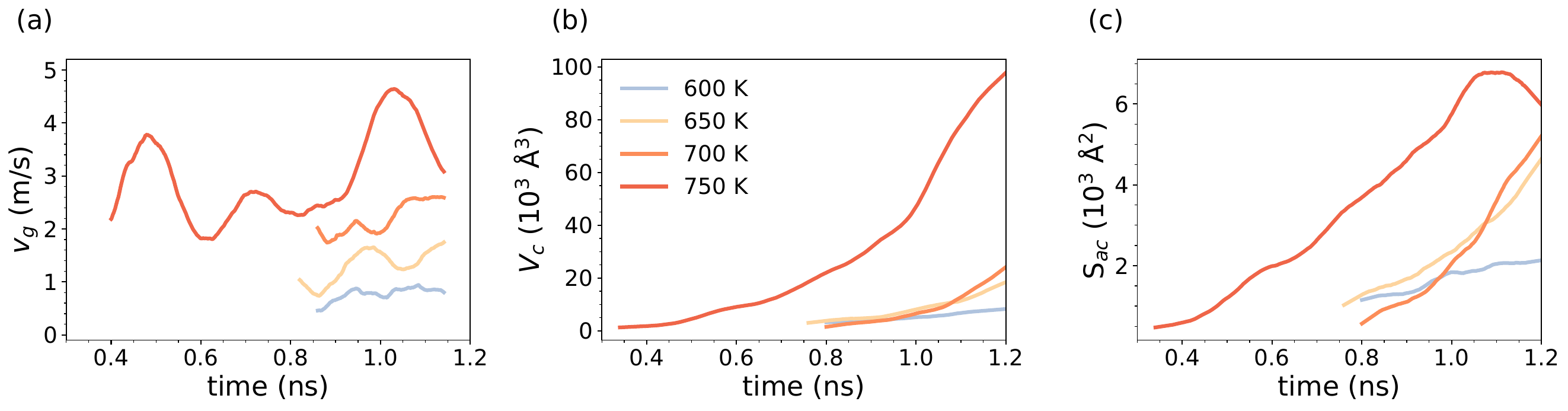}
    \caption{(a) Instantaneous crystal growth velocity $v_g$, (b) volume occupied by the crystalline atoms $V_c$ and (c) area of the crystal-amorphous interface $S_{ac}$ as a function of time at the different temperatures for the SL-LD' model.}

\end{figure}

\begin{figure}[h]
  \renewcommand\figurename{Figure~S$\!\!$}
    \centering
\includegraphics[width=\textwidth,keepaspectratio]{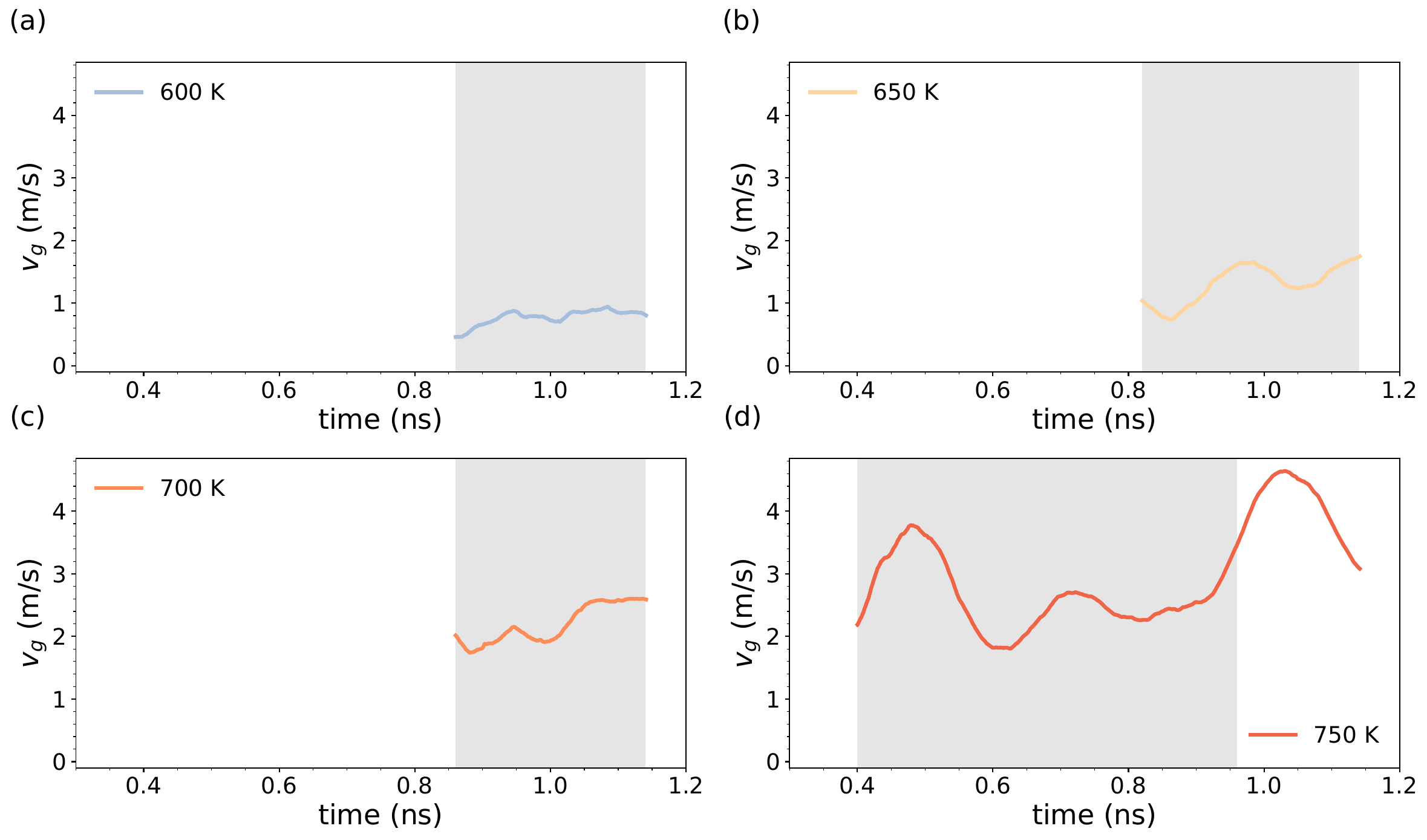}
    \caption{a)-c) Instantaneous crystal growth velocity ($v_g$) as a function of time  at different temperatures for the SL-LD' model. The region highlighted in gray corresponds to the time interval over which we estimated the average crystal growth velocities reported in Table 3 in the article.}
 \label{fig:my_label}
\end{figure}

\begin{figure}[h]
  \renewcommand\figurename{Figure~S$\!\!$}
    \centering
\includegraphics[width=\textwidth,keepaspectratio]{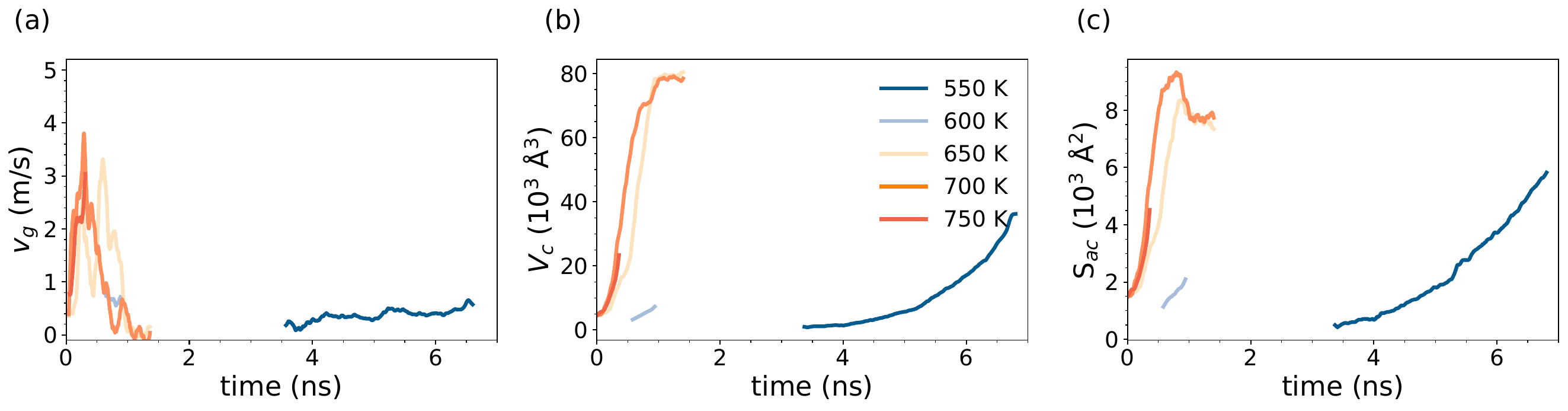}
    \caption{(a) Instantaneous crystal growth velocity $v_g$, (b) volume occupied by the crystalline atoms $V_c$ and (c) area of the crystal-amorphous interface $S_{ac}$ as a function of time at the different temperatures for the bulk simulations at low density (bulk-LD).}

\end{figure}

\begin{figure}[h]
  \renewcommand\figurename{Figure~S$\!\!$}
    \centering
\includegraphics[width=\textwidth,keepaspectratio]{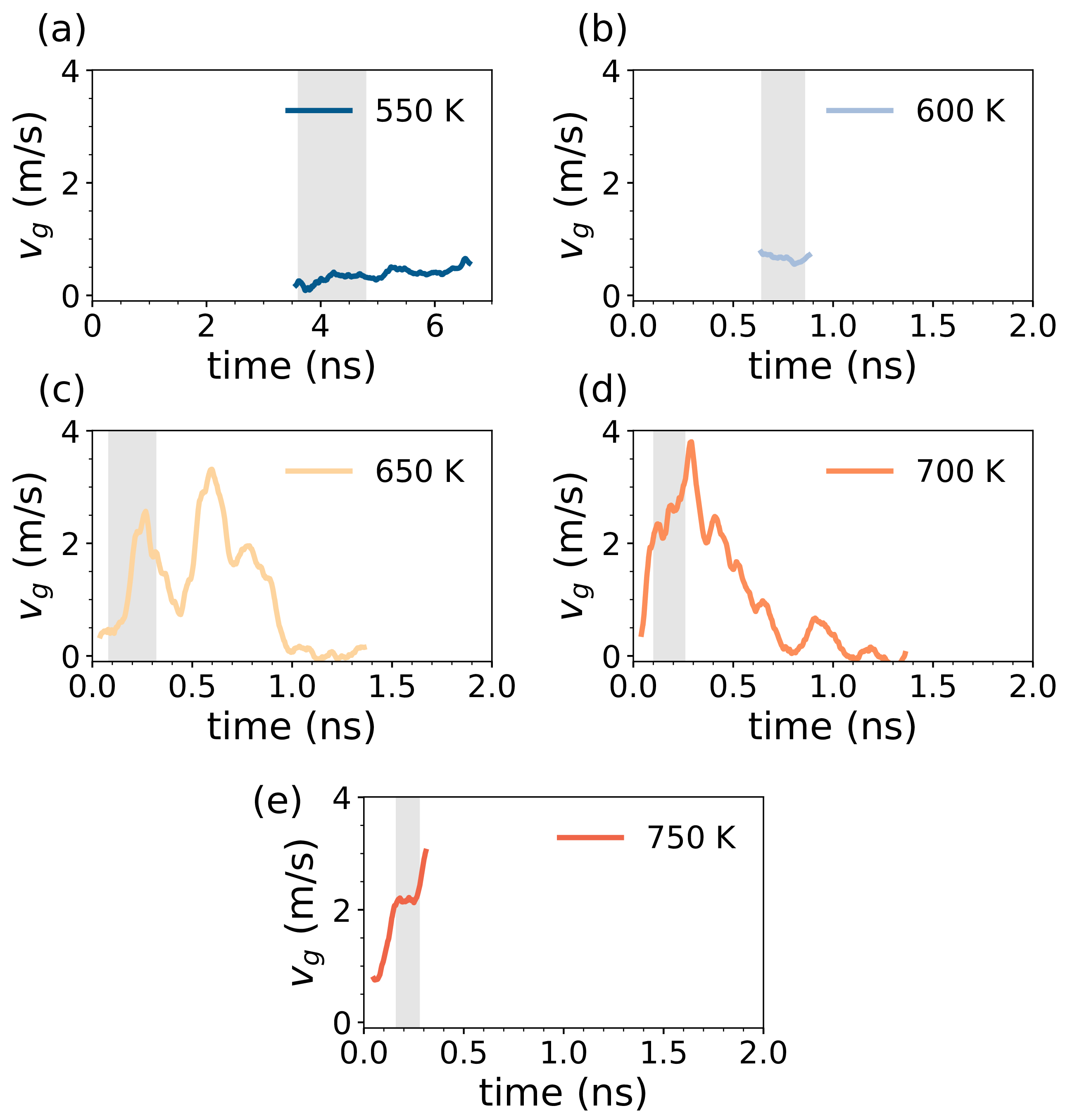}
    \caption{a)-d) Instantaneous crystal growth velocity ($v_g$) as a function of time  at different temperatures for the bulk at low density (bulk-LD). The region highlighted in gray corresponds to the time interval over which we estimated the average crystal growth velocities reported in Table 3 in the article.}

\end{figure}

\begin{figure}[h]
  \renewcommand\figurename{Figure~S$\!\!$}
    \centering
\includegraphics[width=\textwidth,keepaspectratio]{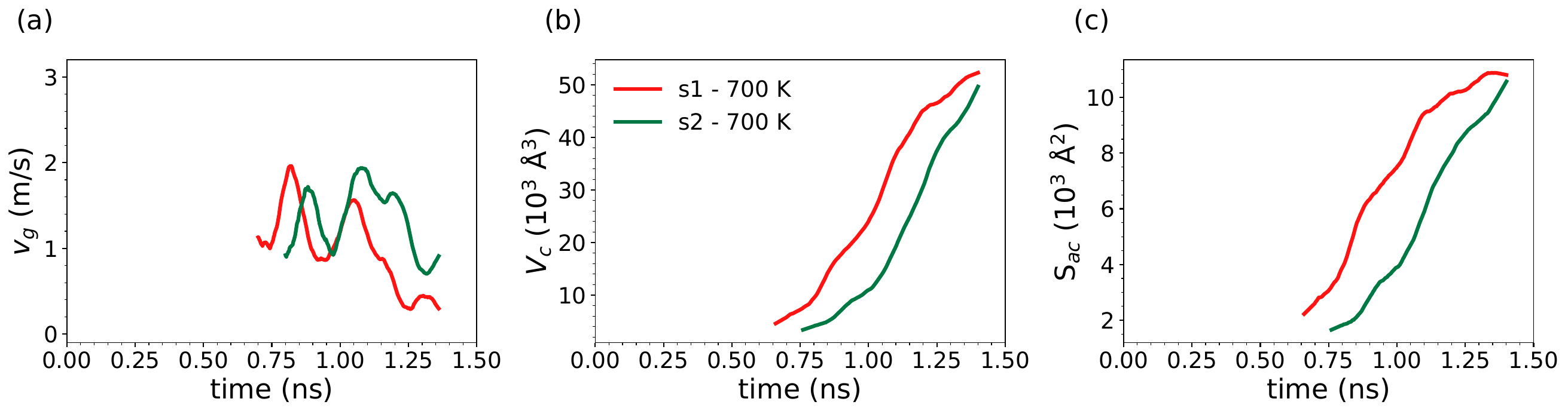}
\includegraphics[width=\textwidth,keepaspectratio]{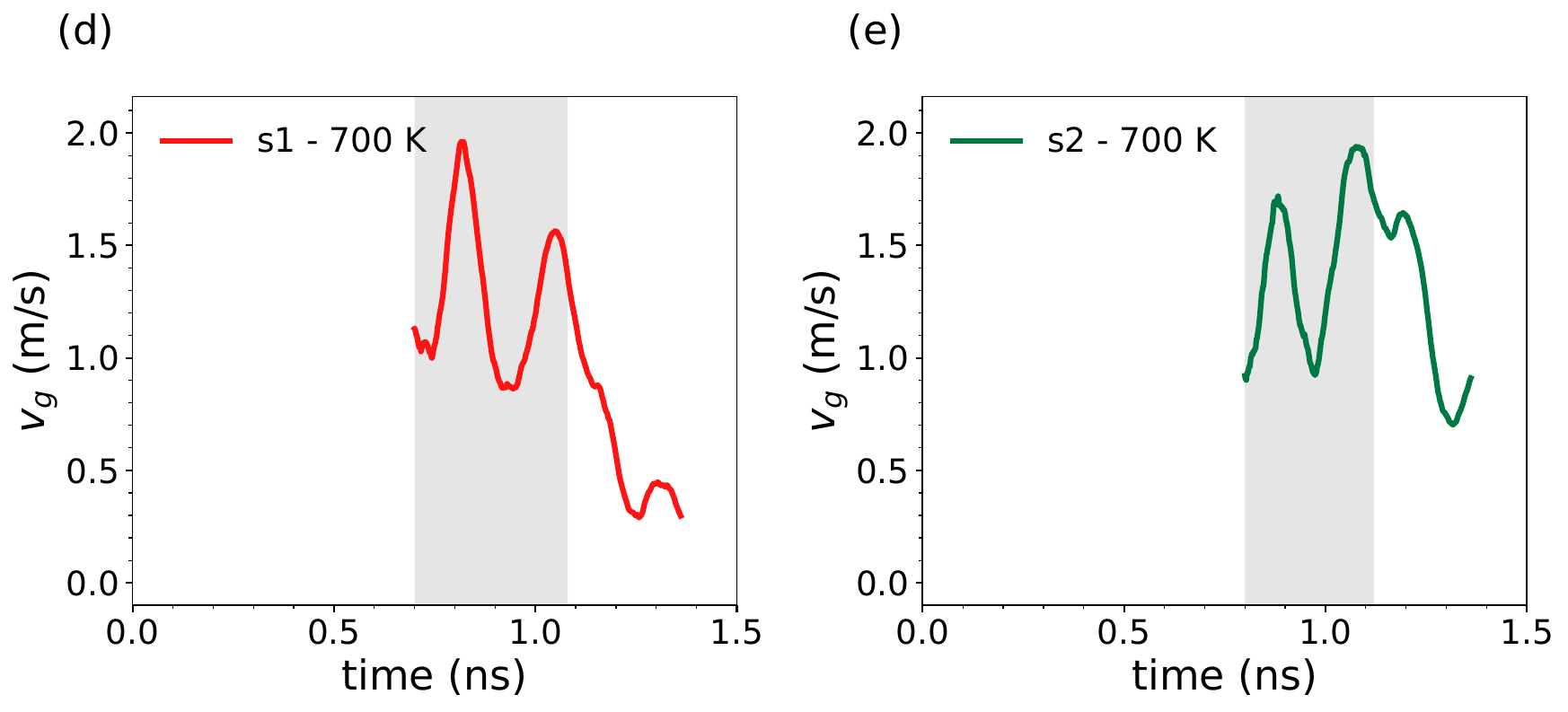}
    \caption{(a) Instantaneous crystal growth velocity $v_g$, (b) volume occupied by the crystalline atoms $V_c$, (c) area of the crystal-amorphous interface $S_{ac}$ as a function of time at 700 K in other two independent models for the SL-HD geometry. (d)-(e)  Instantaneous crystal growth velocity ($v_g$) as a function of time where region highlighted in gray corresponds to the time interval over which we estimated the average crystal growth velocities reported in Table 3 in the article. }

\end{figure}

\begin{figure}[h]
  \renewcommand\figurename{Figure~S$\!\!$}
    \centering
\includegraphics[width=\textwidth,keepaspectratio]{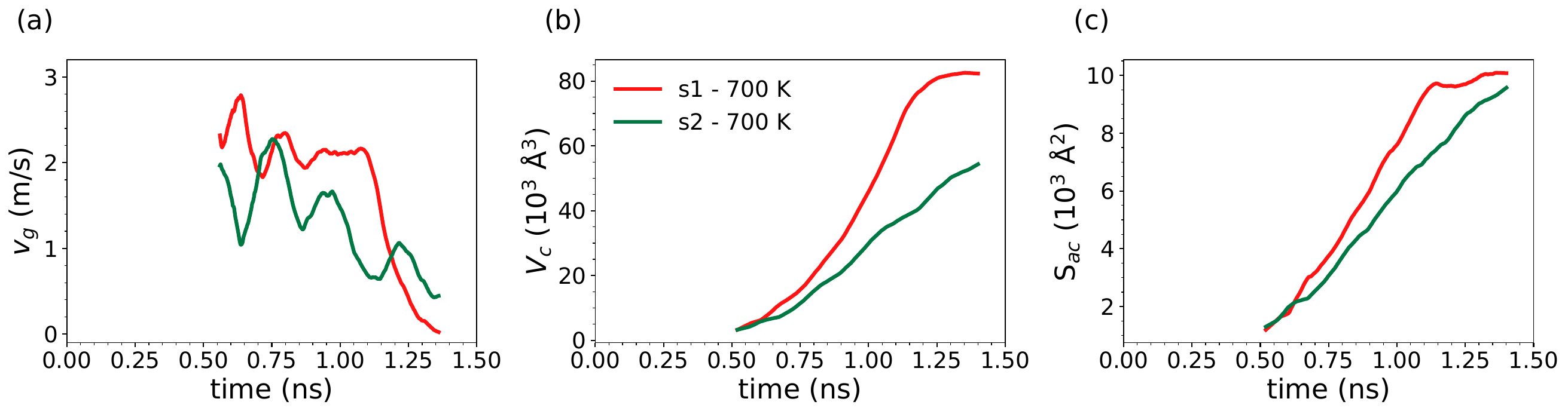}
\includegraphics[width=\textwidth,keepaspectratio]{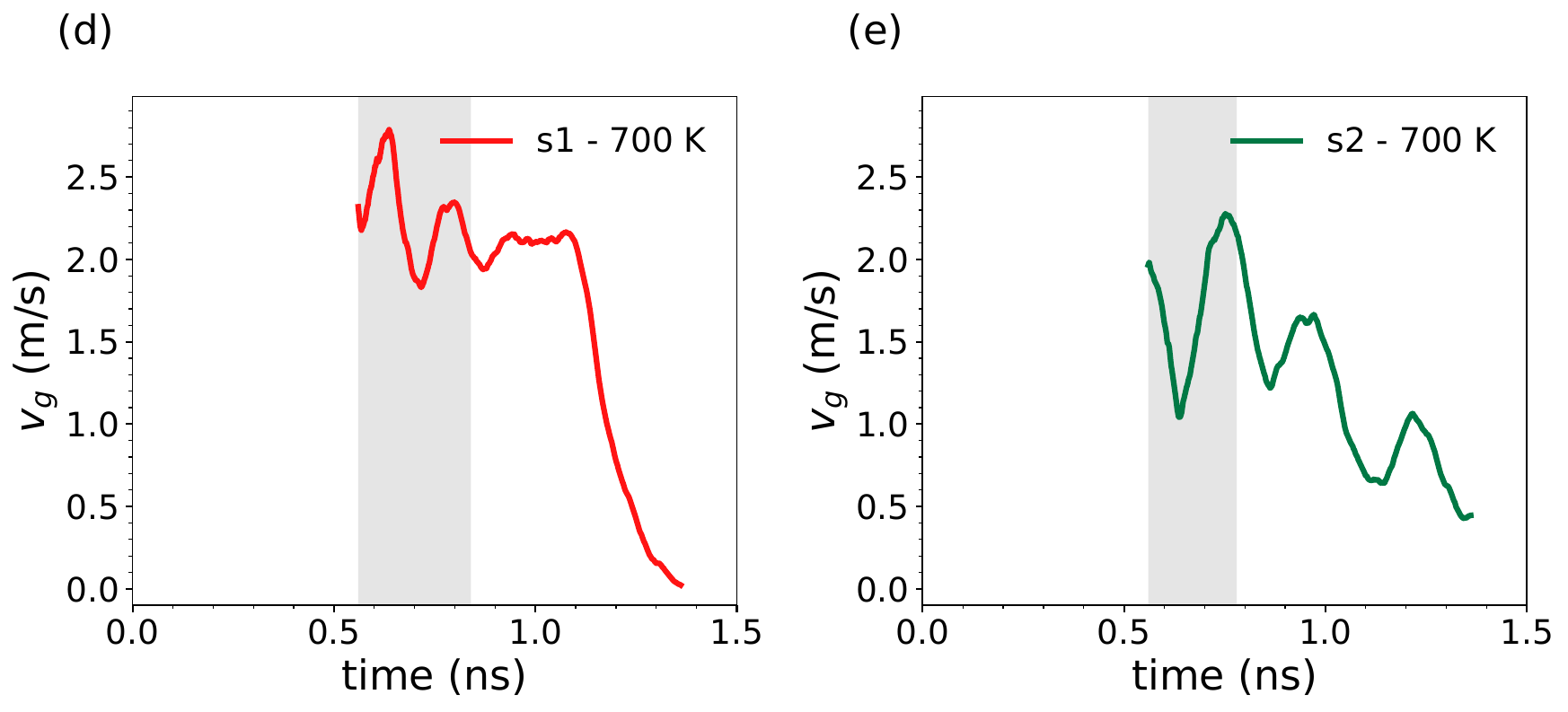}
    \caption{(a) Instantaneous crystal growth velocity $v_g$, (b) volume occupied by the crystalline atoms $V_c$, (c) area of the crystal-amorphous interface $S_{ac}$ as a function of time at 700 K in other two independent models of the SL-LD' geometry. (d)-(e)  Instantaneous crystal growth velocity ($v_g$) as a function of time where region highlighted in gray corresponds to the time interval over which we estimated the average crystal growth velocities reported in Table 3 in the article. }

\end{figure}

\clearpage
\begin{table}[h]
\renewcommand\tablename{Table~S$\!\!$}
\caption{Two-dimensional diffusion coefficient $D$ as a function of time of the SL at the equilibrium density of the hexagonal phase (SL-HD, see article), from NVE simulations at the average temperatures given in the first column. 
We computed $D$ from the two dimensional mean square displacement (MSD) in the plane perpendicular to the slab  thickness in the SL as $<x^2>+<y^2>=4 Dt$. The diffusion coefficient in the slab is compared to those in the bulk at the same density (bulk-HD, see article) at the average temperatures given in the third column. The calculations refer to amorphous models equilibrated at 300 K and then heated and equilibrated at the target temperature in 100 ps. D was then computed  in the subsequent NVE simulations lasting 400 ps.}
\centering
\begin{tabular}{ccccc}
\hline
\hline
Temperature (K)& $D$  (10$^{-6}$cm$^2$/s)  & Temperature (K) & $D$  (10$^{-6}$cm$^2$/s)\\
& SL-HD & & Bulk-HD \\
\hline
517&0.22& 501 & 0.13\\
560 &0.42& 559 & 0.40\\
603 &1.07& 601 & 1.04\\
651&2.44&636& 2.08\\
702 &4.90& 687 & 4.69\\
756 &8.99& 755 & 9.13\\
\hline
\hline
\end{tabular}
\label{tab:svelocities}
\end{table}
\clearpage
\begin{figure}[h]
\renewcommand\figurename{Figure~S$\!\!$}
    \centering
{\includegraphics[height=8cm]{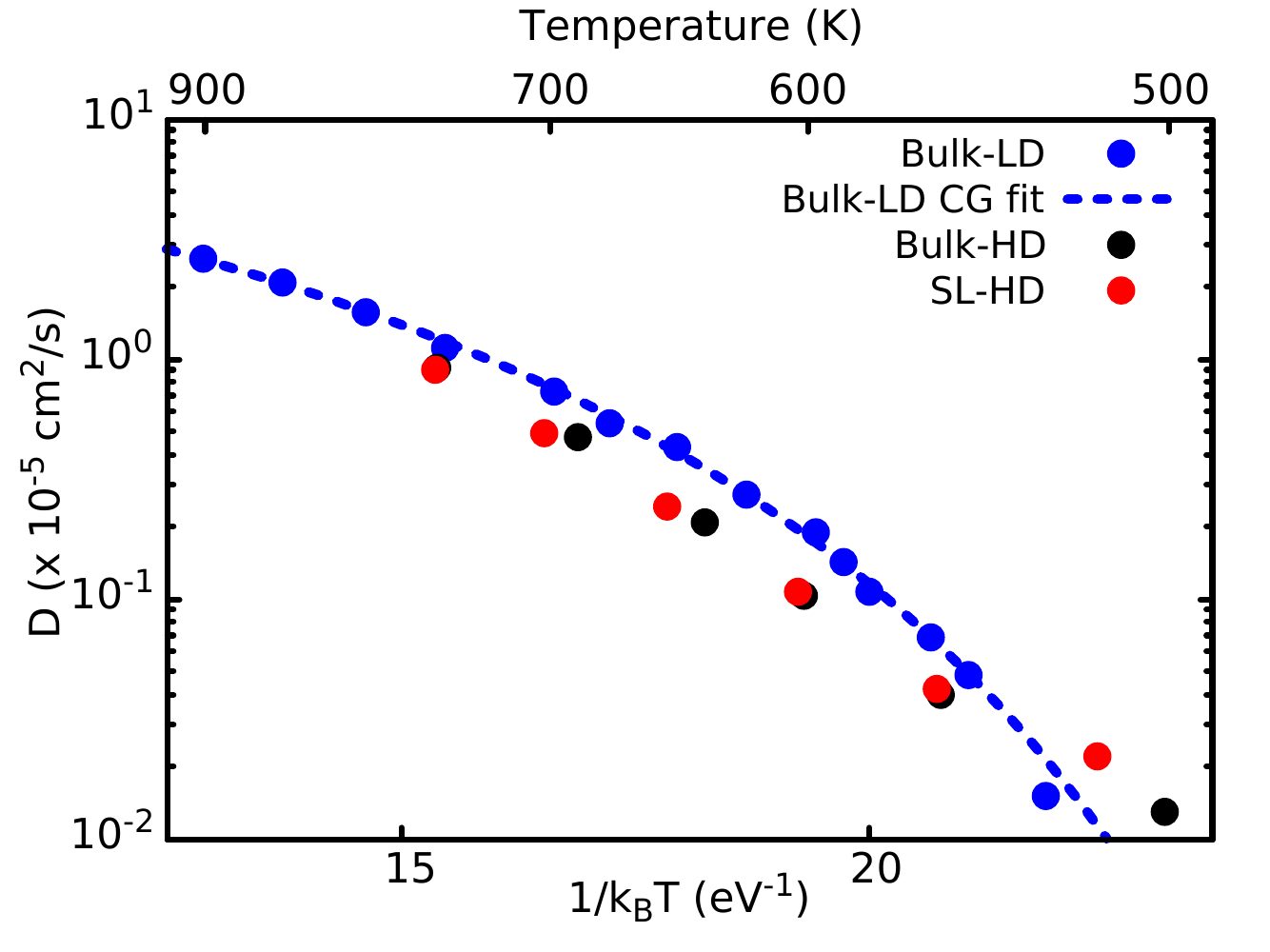} } 
\caption{Diffusion coefficient $D$ (10$^{-5}$ cm$^2$/s) as a function of temperature from bulk NN simulations at the experimental density of the amorphous phase (bulk-LD) and of the crystalline hexagonal phase (bulk-HD) and of the superlattice model SL-HD (see article). }
\label{sdiffusion}
\end{figure}
\newpage

\begin{figure}[h]
  \renewcommand\figurename{Figure~S$\!\!$}
    \centering
\includegraphics[width=\textwidth,keepaspectratio]{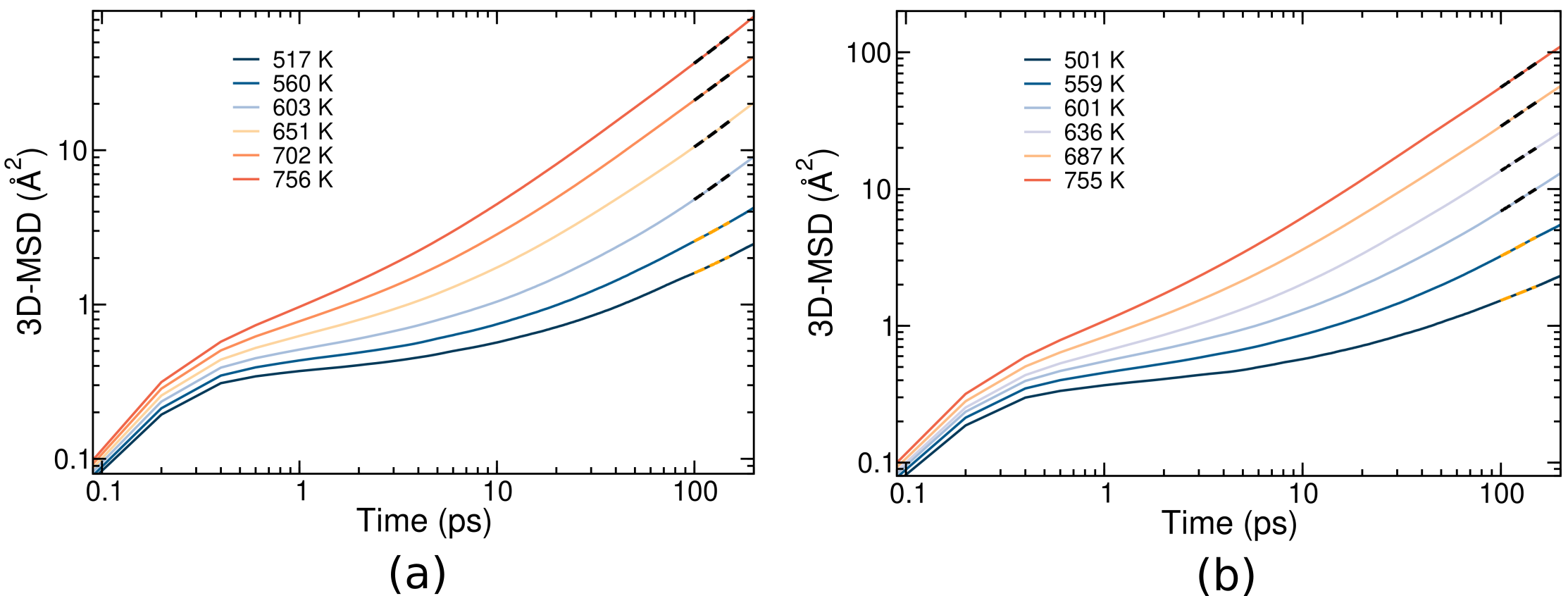}
    \caption{Mean square displacement (MSD) as a function of time from NVE simulations at the average temperatures given in the inset for a) SL at the equilibrium density of the hexagonal phase (SL-HD) and b) for the bulk at the same density (bulk-HD, see article).
      For the sake of comparison with the bulk, the 3D-MSD is plotted for the SL as well, where 3D-MSD= 3/2($<x^2>$ + $<y^2>$).}

\end{figure}

\end{document}